\documentclass[aps,prb,showpacs,twocolumn,superscriptaddress,floatfix]{revtex4-2}
\usepackage{amsmath,amssymb,mathtools}

\usepackage{graphicx}
\usepackage{xcolor} 
\usepackage{dcolumn}
\usepackage{bm}
\usepackage{hyperref}
\usepackage{comment}

\begin{document}

\preprint{APS/123-QED}

\title{Spin-basis wavefunctions for the one-dimensional Kitaev model}

\author{Alwyn Jose Raja}
\email{alwynjoseraja2000@gmail.com}
 
\author{Rajesh Narayanan}
\email{rnarayanan@zmail.iitm.ac.in}

\affiliation{
 Department of Physics, Indian Institute of Technology Madras, Chennai 600036, India}

\author{R. Ganesh}
\email{r.ganesh@brocku.ca}
\affiliation{
Department of Physics, Brock University, St. Catharines, Ontario L2S 3A1, Canada
}

\date{\today}

\begin{abstract}
Magnetic phases with quantum entanglement are often expressed in terms of parton wavefunctions. Relatively few examples are known where wavefunctions can be directly written down in the spin basis. In this article, we consider the spin-$S$ Kitaev model in one dimension. For $S=1/2$, its eigenstates can be written using a Jordan-Wigner fermionic representation. Here, we present ground state wavefunctions for any $S$ directly in the spin basis. 
The states we propose are valence bond arrangements, with bonds having singlet or triplet character for $S=1/2$. For $S>1/2$, we use bond-states that serve as analogues of singlets and triplets.  
We establish the validity of our wavefunctions using a perturbative approach starting from an anisotropic limit. 
For half-integer $S$ and periodic boundaries, we have exponential ground state degeneracy. The ground states are subject to a non-local constraint.
They have `triplets' superposed on a background of singlets, but with the total number of triplets constrained to be even.
For integer $S$, a unique ground state emerges, composed purely of `triplets'. Our spin-basis wavefunctions, while not exact, capture the dominant weight of the ground state(s).
We obtain good agreement against exact diagonalization wavefunctions and Jordan-Wigner spectra. 
\end{abstract}


\maketitle

\section{Introduction}
Quantum magnets are a fertile playground for entanglement and topology\cite{fazekas1999,auerbach2012,kitaev2006anyons}. To explore these ideas, wavefunctions are typically written down by invoking fractionalization. The resonating valence bond (RVB) phase of Anderson and coworkers is a case in point\cite{anderson1987resonating,liang1988some,zhou2017quantum}. It proposes a wavefunction in the spin basis, consisting of a superposition of singlet covers. In practice, the RVB wavefunction is written in terms of fractionalized bosonic or fermionic operators\cite{Lee2006}. 
A second cornerstone is the Affleck, Kennedy, Lieb and Tasaki (AKLT) state\cite{affleck1988valence}, expressed in terms of fractionalized spin-$1/2$ degrees of freedom. 
A third example is the 1D XY model, solved using Jordan Wigner fermionization\cite{Lieb1961}. The ground state is expressed as that of a sea of non-interacting fermions. 
In all three examples, the wavefunction, even if it were known exactly, cannot be easily expressed in terms of the original spin operators.
In contrast, the Majumdar-Ghosh\cite{majumdar1969next} and Shastry-Sutherland\cite{shastry1988exact} models provide exact solutions in terms of the original spin operators.

In this article, we present wavefunctions for the Kitaev chain directly in terms of the constituent spins. This model has been extensively studied as the one-dimensional version of Kitaev's exactly solvable model on the honeycomb lattice\cite{kitaev2006anyons}. It can also be viewed as the one-dimensional compass model\cite{Brzezicki2007,nussinov2015compass}. Its frustrated character can be seen in the classical $S\rightarrow \infty$ limit, which yields high ground state degeneracy. Classical ground states can be constructed  starting from `cartesian states', where every alternating bond is maximally satisfied\cite{baskaran2008spin,Khatua2021}. Additional ground states appear, which can be viewed as pathways that connect pairs of cartesian states.
Below, we demonstrate that the quantum version of the model also yields high degeneracy as long as the spin quantum number, $S$, is a half-integer.

We build upon the approach of Gordon and Kee\cite{gordon2022insights} who developed a perturbative approach to describe the ground state(s) of this model. We
propose wavefunctions for the spin-$S$ model. They are not exact solutions, rather they are  approximations that carry a significant weight of the true ground state(s). Two interesting features emerge: (i) There is a qualitative difference between integer and half-integer values of $S$. This is reminiscent of the  Heisenberg chain, where such a difference originates from a topological term\cite{Tasaki2020}. 
(ii) For half-integer $S$, ground state wavefunctions obey a global constraint. They are obtained as dimer covers, where each dimer is assigned a `singlet' or a `triplet' wavefunction. (For $S>1/2$, we use suitably defined analogues of the $S=1/2$ singlet and triplet states). The global constraint forces the total number of triplets to be even. We argue that this is a `global effect' that cannot be altered by local perturbations.

We describe the quantum spin-$S$ problem using various approaches below. We define the model in Sec.~\ref{sec.model}.
In Sec.~\ref{sec.ED}, we discuss exact diagonalization results for small system sizes. Sec.~\ref{sec.JW} presents a Jordan-Wigner treatment for $S=1/2$. In Sec.~\ref{sec.spinbasis}, we propose spin-basis wavefunctions for ground states. In Sec.~\ref{sec.pert}, we justify the form of wavefunctions using a perturbative approach starting from the anisotropic $K_x \gg K_y$ limit. In Sec.~\ref{sec.isotropic}, we discuss the isotropic $K_x=K_y$ point.
We conclude with a discussion in Sec.~\ref{sec.discussion}.

\section{Model: Spin-$S$ Kitaev Chain}
\label{sec.model}
We consider a one-dimensional spin system with a two-site unit cell and a spin-$S$ moment at each site. We have $N$ spins that are coupled by nearest-neighbour bonds that alternate between $X$ and $Y$ character. The system is described by the Hamiltonian
\begin{equation}
    H= K_x \sum_{i=1}^{N/2}  S_{A_i}^xS_{B_i}^x + K_y \sum_{i=1}^{N/2} S_{B_i}^y S_{A_{i+1}}^y.
    \label{eq.Hamiltonian}
\end{equation}
Here $A$ and $B$ label the two sites within the unit cell. The $X$ bonds are intra unit cell couplings while the $Y$ bonds are inter unit cell couplings. We may enforce periodic boundary conditions by identifying the $i^\mathrm{th}$ and $(i+N/2)^\mathrm{th}$ unit cells. We assume that $N$ is even and $N\geq 4$, so as to allow for alternating $X$ and $Y$ bonds. For concreteness, we consider $K_x$ and $K_y$ to be positive coupling constants throughout this manuscript. Our results can be easily adapted to cases where one or both are negative, using basis changes generated by local rotations. 

\begin{figure}
\includegraphics[width=\columnwidth]{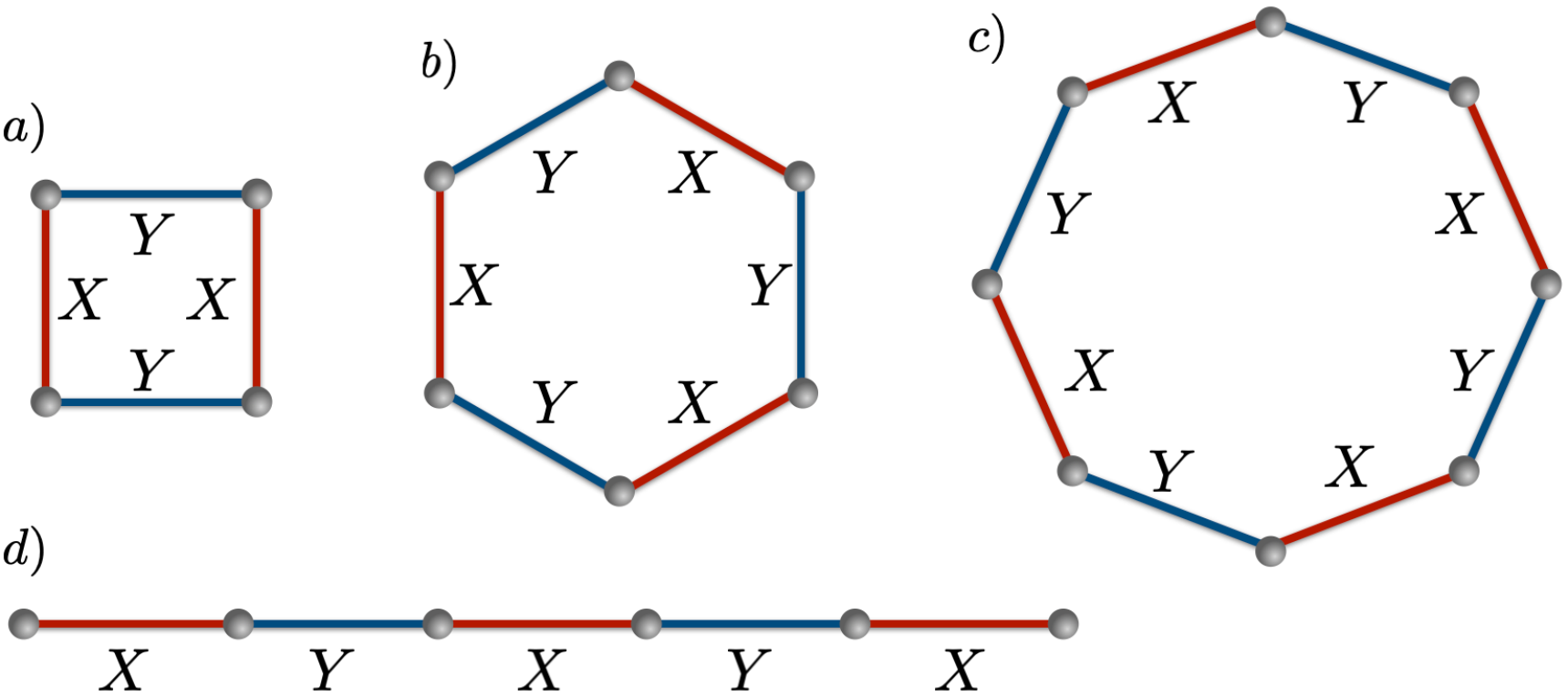}
\caption{Kitaev chains with periodic boundary conditions for (a) $N=4$, (b) $N=6$ and (c) $N=8$. With open boundary conditions, we fix the bonds at the ends to be of the X type, for concreteness. This is shown for $N=6$ in (d). }
\label{fig.clusters}
\end{figure}

This model describes a much wider class of systems than appears at first glance. As shown in Ref.~\cite{sen2010spin}, the isotropic limit ($K_x=K_y=1$) of this model can be rewritten as
\begin{equation}
    H=\sum_i^N S_i^xS_{i+1}^y.
\end{equation}
It can also be mapped to a 3-unit cell `Kitaev-XYZ' chain with a repeating sequence of $X-Y-Z$ bonds \cite{yang2025emergent}, 
\begin{equation}
    H=  \sum_i^\frac{N}{3}S_{A_i}^xS_{B_i}^x + \sum_i^\frac{N}{3}S_{B_i}^y S_{C_{i}}^y + \sum_i^\frac{N}{3} S_{C_i}^z S_{A_{i+1}}^z.
\end{equation}
In Appendix~\ref{app.Barlow}, we  demonstrate that this Hamiltonian describes any chain with bonds of $X$, $Y$ and $Z$ types, with a constraint that two adjacent bonds cannot be of the same type. These chains are analogous to the H\"agg code representation of close-packed structures\cite{Hagg1943,KrishnaPandey1981}. 

To describe eigenstates of the Hamiltonian in Eq.~\ref{eq.Hamiltonian}, we may use conserved quantities defined on bonds\cite{sen2010spin},
\begin{eqnarray}
    W_{i,X} = e^{i\pi S_{A_i}^y}e^{i\pi  S_{B_i}^y},~~    W_{i,Y} = e^{i\pi S_{B_i}^x}e^{i\pi S_{A_{i+1}}^x},
    \label{eq.Ws}
\end{eqnarray}
defined on X and Y bonds respectively. All $W$'s square to unity and commute with the Hamiltonian. 
For integer values of $S$, the $W$'s commute with one another. With periodic boundaries, this leads to $N$ independent conserved quantities. 
For half-integer values of $S$, $W$'s on neighbouring bonds do not commute with one another. This leads to $N/2$ conserved quantities, say corresponding to $W$'s on the X bonds alone. In both cases, these conserved quantities can be viewed as fragmenting the Hilbert space \cite{mohapatra2023pronounced,you2022quantum}. In our discussion below, we describe ground state wavefunctions and characterize them as lying in specific $W$-sectors.

\section{Exact Diagonalization (ED)}
\label{sec.ED}

\begin{figure*}
    \includegraphics[width=0.246\linewidth]{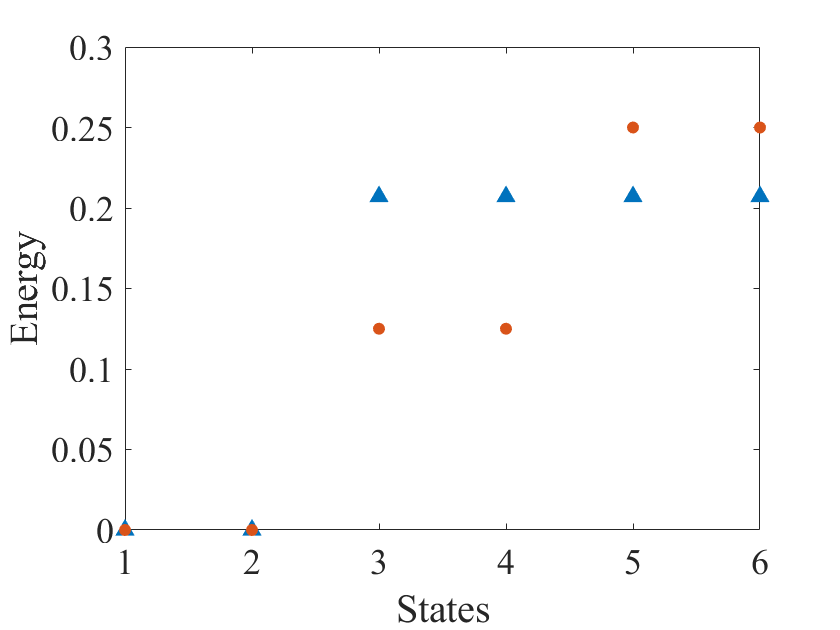}
    \includegraphics[width=0.246\linewidth]{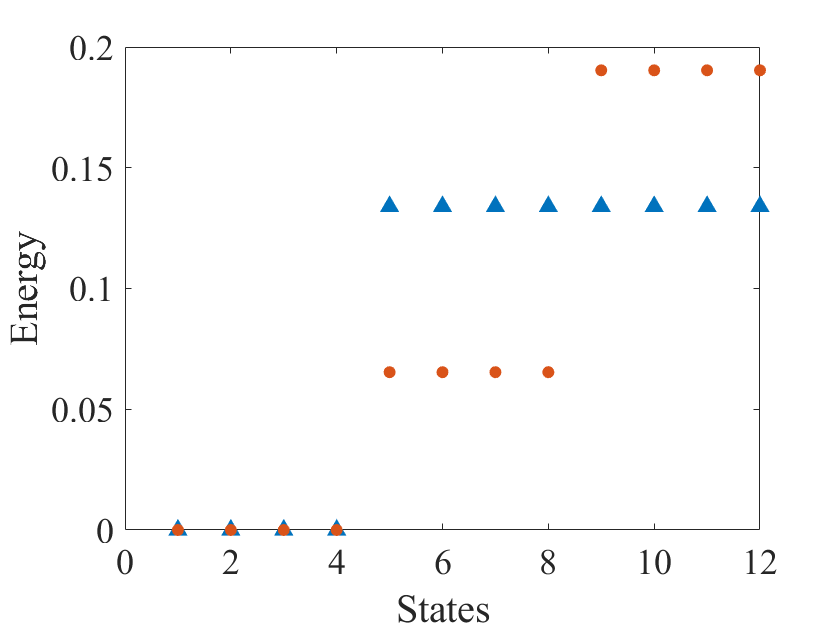}
    \includegraphics[width=0.246\linewidth]{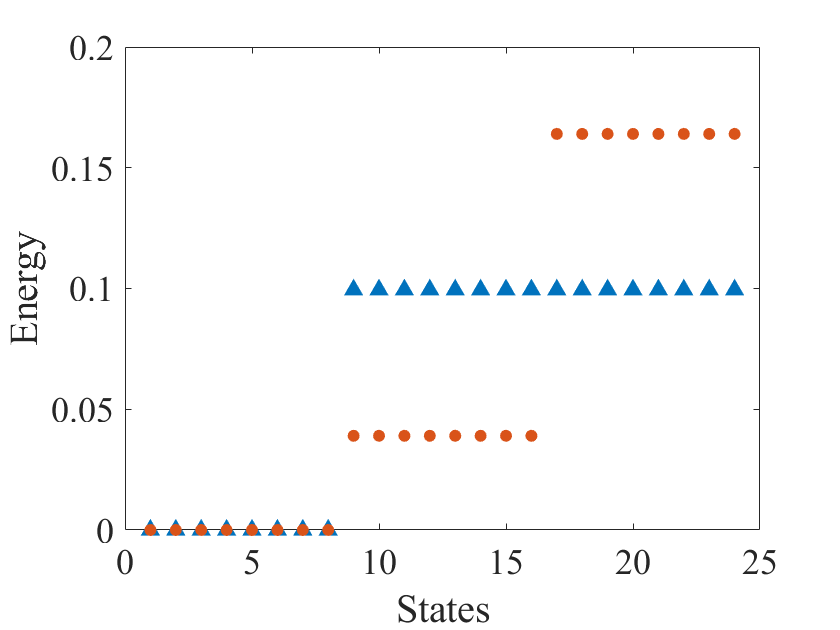}
    \includegraphics[width=0.246\linewidth]{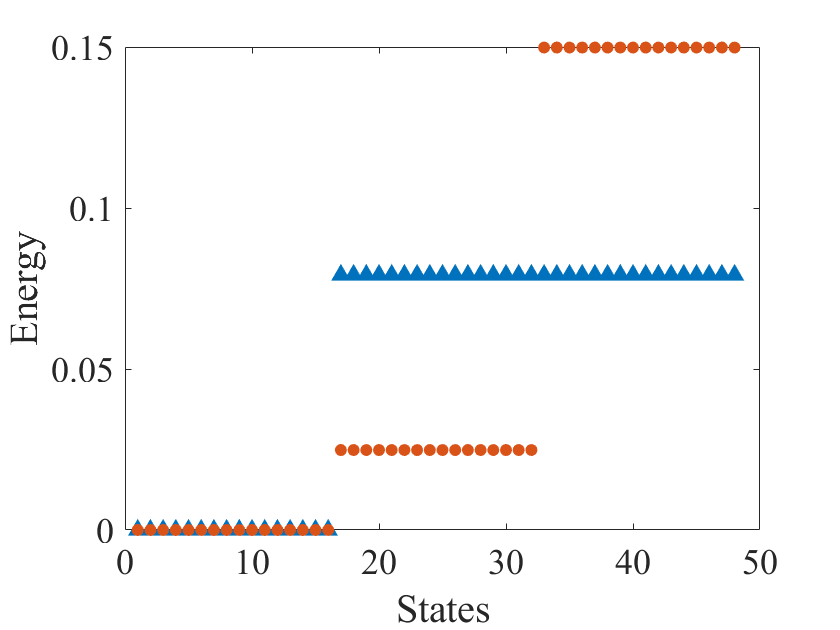}    
    \caption{Low-energy spectrum for $S=\frac{1}{2}$ PBC chain for $N=4,6,8,10$ (from left to right). We only show the lowest few states for clarity. In each case, the ground state has a degeneracy of $2^{\frac{N}{2}-1}$. We fix $K_x=1$. Blue triangles correspond to $K_y=1$, while red circles correspond to $K_y=0.75$. The first excited state is $2^\frac{N}{2}$-fold degenerate at the isotropic point, splitting into two degenerate sets with anisotropy. 
    Analytic arguments for these patterns are given in Secs.~\ref{sec.pert} and \ref{sec.isotropic} below.
    Energy eigenvalues have been shifted to fix the ground state energy at zero. }
    \label{fig. Spin half ED PBC}
\end{figure*}
\begin{figure*}
    \includegraphics[width=0.246\linewidth]{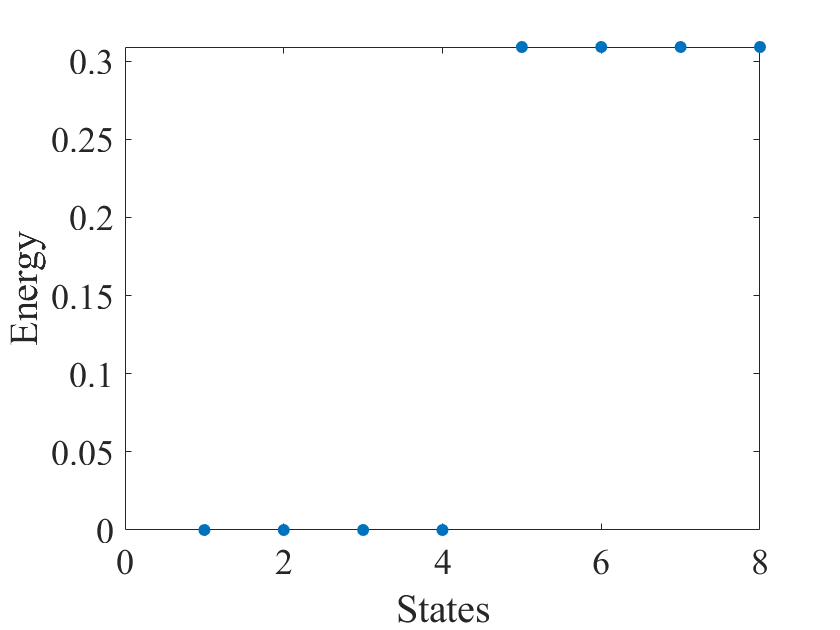}
    \includegraphics[width=0.246\linewidth]{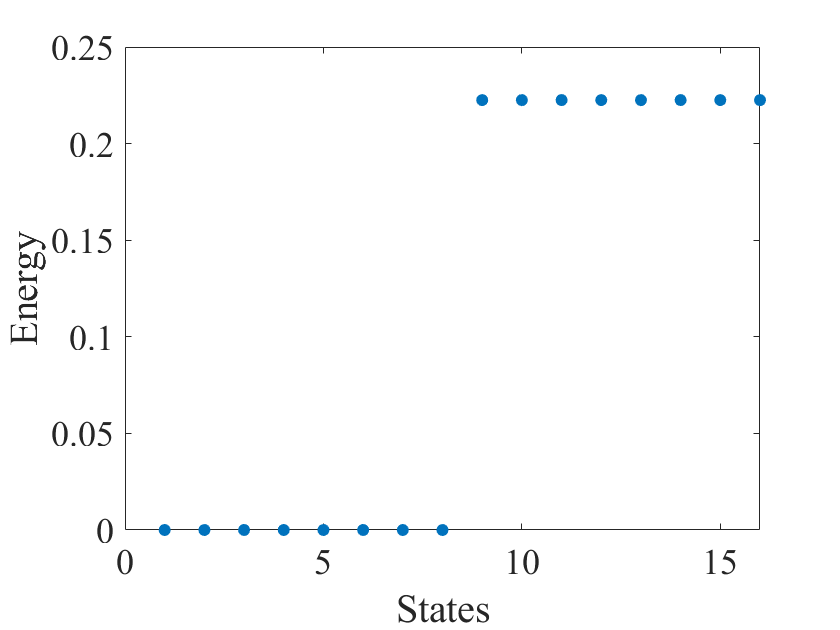}
    \includegraphics[width=0.246\linewidth]{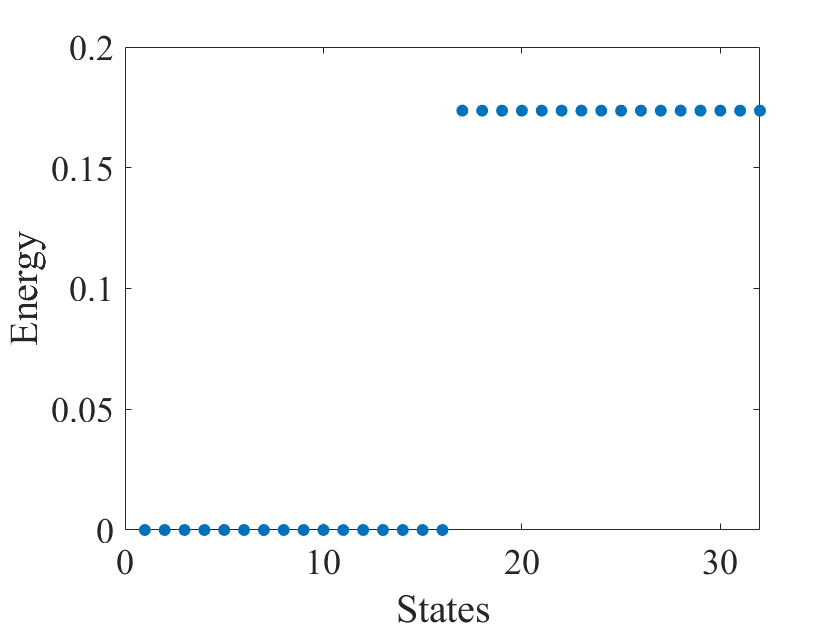}
    \includegraphics[width=0.246\linewidth]{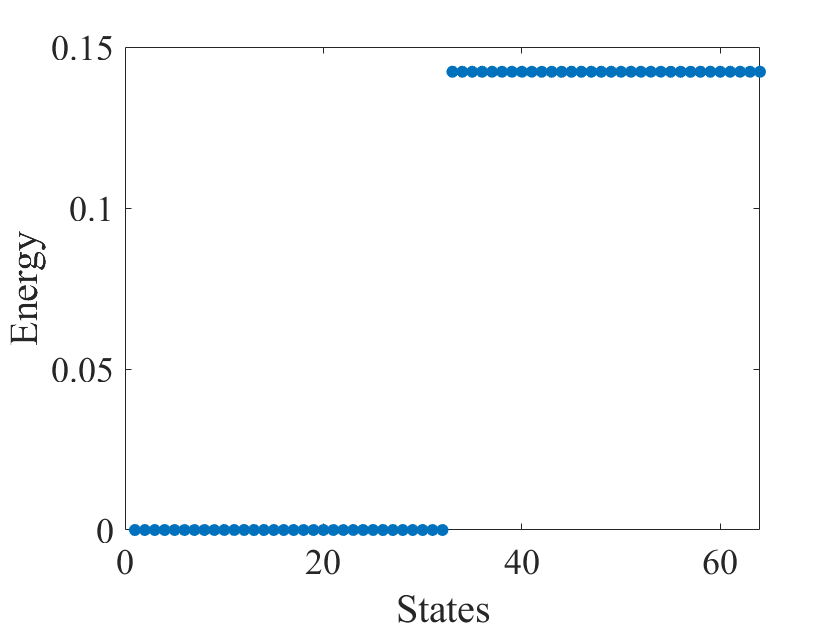}   
    \caption{Low-energy spectrum for $S=\frac{1}{2}$ OBC chain for $N=4,6,8,10$ (from left to right). The ground state and the first excted state are $2^{\frac{N}{2}}$-fold degenerate. We show data for $K_x=K_y=1$. There is no qualitative difference for $K_x\neq K_y$, as long as both couplings are non-zero. Energies have been shifted to set the ground state energy to zero.}
    \label{fig. Spin half ED OBC}
\end{figure*}
\begin{figure*}
    
    \includegraphics[width=0.45\linewidth]{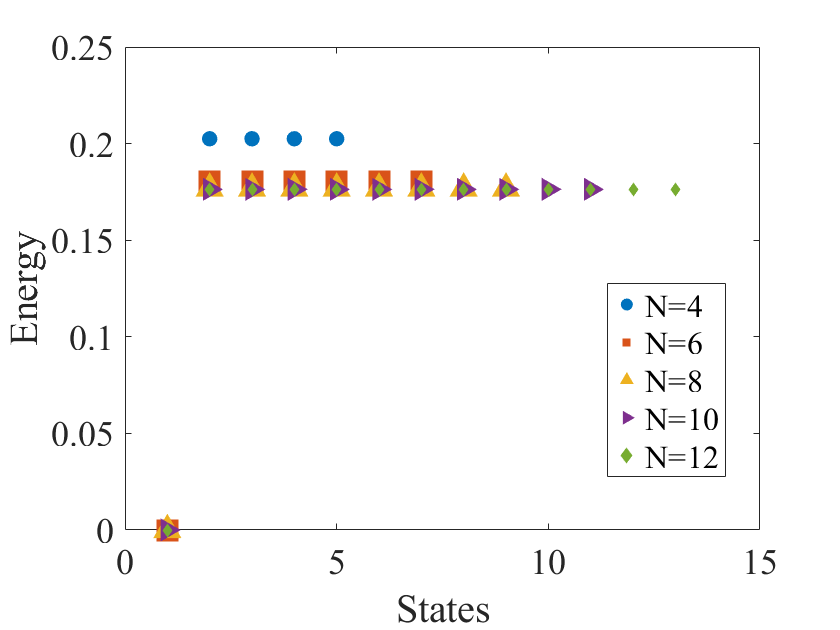}
    \includegraphics[width=0.45\linewidth]{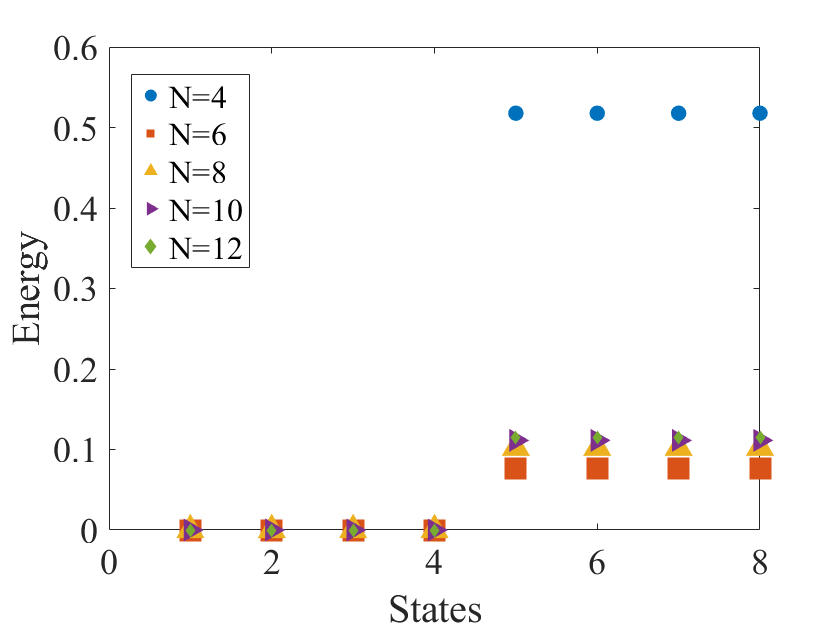}
    \caption{Low-energy spectrum for $S=1$ with $K_x=K_y$ for PBC (left) and OBC (right). Energies are obtained from exact diagonalisation for system sizes $N=4$ to $N=12$. PBC leads to a unique ground state and an $N$-fold degenerate first excited state. With OBC, we only show the lowest eight energies. The ground state is always four-fold degenerate. Y-axis values have been shifted to set the ground state energy to zero. }
    \label{fig. Spin 1 ED PBC and OBC}
\end{figure*}
For small system sizes, the ground state(s) of the Hamiltonian in Eq.~\ref{eq.Hamiltonian} can be determined by exact diagonalization.  
We treat the system size $N$ as a tuning parameter, considering both open (OBC) and periodic boundary conditions (PBC). 
We discuss certain robust patterns observed in exact diagonalization spectra below. In Secs.~\ref{sec.spinbasis}-\ref{sec.isotropic}, we will present analytic arguments demonstrating that these patterns hold for all system sizes -- with a qualitative difference between integer and half-integer values of $S$.

\noindent With periodic boundaries and integer values of $S$:
\begin{itemize}
\item 
We have a unique ground state for any $N$ and any non-zero values of $K_x$, $K_y$.

\end{itemize}
With periodic boundaries and half-integer values of $S$:
\begin{itemize}
\item The ground state is highly degenerate. The degeneracy grows exponentially with $N$, as $2^{N/2-1}$. 
\item For $K_x \neq K_y$, the first excited state also has a degeneracy of $2^{N/2-1}$. The gap to the first excited state decreases with system size. In the thermodynamic $N\rightarrow\infty$ limit, the gap closes as the ground state and first excited state become degenerate. We surmise that the ground state degeneracy in the thermodynamic limit is at least $2^{N/2}$.
\item For the isotropic $K_x = K_y$ point, we have a higher first-excited-state-degeneracy of $2^{N/2}$. The gap to the first excited state decreases with system size and vanishes in the thermodynamic limit. For $N\rightarrow \infty$, the ground state degeneracy has a lower bound of $3\times 2^{N/2-1}$.
\end{itemize}

With open boundaries (assuming, for concreteness, that both edges are of the same type, e.g., both are X bonds):
\begin{itemize}
\item For integer-$S$, we find a ground state degeneracy of four. 
\item With half-integer $S$, the ground state has a degeneracy of $2^{N/2}$ for any finite $N$. 
\end{itemize}
We show exact diagonalization spectra that are in line with these assertions in Figs.~\ref{fig. Spin half ED PBC}, ~\ref{fig. Spin half ED OBC} and ~\ref{fig. Spin 1 ED PBC and OBC}. We present data for $S=\frac{1}{2}$ and $S=1$ as representative cases for half-integer- and integer-$S$. The same qualitative behaviour is seen in ED results for higher values of spin as well, as shown in Appendix D for $S=\frac{3}{2},2,\frac{5}{2}$.

\section{Jordan-Wigner Fermionization}
\label{sec.JW}

We now present exact solutions for the $S=1/2$ case following the well-known Jordan-Wigner prescription. 
We first present the general framework before distinguishing between periodic and open boundary conditions. 
For the 2-unit cell Kitaev spin chain, we define parton operators using
\begin{eqnarray}
    S^+_{A_i}=f_{A_i}^\dagger\prod_{j<i}(1-2~\hat{n}_{A_j})\prod_{k<i}(1-2~\hat{n}_{B_k}), \nonumber \\
    S^+_{B_i}=f_{B_i}^\dagger\prod_{j\leq i}(1-2~\hat{n}_{A_j})\prod_{k<i}(1-2~\hat{n}_{B_k}),
    \label{eq.JWdefs}
\end{eqnarray}
where $f_{{A/B}_i}$ is a fermionic annihilation operator at site $(i,A/B)$ and $\hat{n}_{{A/B}_i} = f_{{A/B}_i}^\dagger f_{{A/B}_i}$ is the number operator. The site indices $i$ and $j$ are counted differently for periodic and open boundary conditions as described below.

\subsection{Open Boundary Conditions (OBC)}

Site indices are counted from one end of the chain, giving rise to the Hamiltonian 
\begin{eqnarray}
 H_{OBC}=\frac{K_x}{4}\sum_{i=1}^\frac{N}{2} (f^\dagger_{A_i}f_{B_i}+f^\dagger_{A_i}f^\dagger_{B_i} + h.c.)+  \nonumber \\
 \frac{K_y}{4}\sum_{i=1}^{\frac{N}{2}-1}  (f^\dagger_{A_{i+1}}f_{B_i}+f^\dagger_{A_{i+1}}f^\dagger_{B_i} + h.c.).
\end{eqnarray}
We define Majorana fermions using 
\begin{eqnarray}
c_{\alpha_i}=f_{\alpha_i}+f^\dagger_{\alpha_i};~~~~
id_{\alpha_i}=f_{\alpha_i}-f^\dagger_{\alpha_i},    
\end{eqnarray}
where $\alpha=A,B$ encodes the sublattice degree of freedom. We have 2 Majorana fermions per site and 4 per unit cell. In terms of these new operators,

\begin{equation}
    H_{OBC}= \Big[\sum_{i=1}^\frac{N}{2} \frac{iK_x}{4}(c_{B_i}d_{A_i}) + \sum_{i=1}^{\frac{N}{2}-1}\frac{iK_y}{4}(c_{B_i}d_{A_{i+1}})\Big].
    \label{eq.Hsingleparticle}
\end{equation}
Remarkably, for any choice of $(K_x,K_y)$, $c_{A_i}$ and $d_{B_i}$ do not appear in the Hamiltonian. They amount to $N$ Majorana zero modes, or equivalently $\frac{N}{2}$ fermionic zero-energy modes. 

The ground state of the spin model is a many-particle state, obtained by filling single-particle states that arise from diagonalizing the Hamiltonian. To minimize energy, all single-particle states with negative (positive) energy are filled (empty). However, single-particle states with zero energy may either be filled or left empty. 
This yields a ground state degeneracy of $2^{N/2}$. The degeneracy remains unchanged as $K_x$ and $K_y$ are varied.

\subsection{Periodic Boundary Conditions (PBC)}

With periodic boundaries, we designate a unit cell at the edge (an arbitrary choice) as the origin. The Hamiltonian acquires an additional term, when compared to the OBC case, given by
 \begin{eqnarray}
\nonumber K_yS_{B_{\frac{N}{2}}}^yS_{A_1}^y =(-1)^{\mathcal{N}+1}\frac{K_y}{4} \times \\
(f^\dagger_{B_{\frac{N}{2}}}f_{A_1}+f^\dagger_{A_1}f_{B_{\frac{N}{2}}}
-f^\dagger_{B_{\frac{N}{2}}}f^\dagger_{A_1}-f_{A_1}f_{B_{\frac{N}{2}}}).~~~~~
\label{eq.per}
\end{eqnarray} \\
Here, $\mathcal{N} = \sum_{j=1}^{N/2} (\hat{n}_{A_j} + \hat{n}_{B_j})$ represents the total number of particles in the system. The factor of $(-1)^{\mathcal{N}+1}$ arises from the Jordan-Wigner `string' that goes around the loop. 
To diagonalize the single-particle Hamiltonian, we 
adopt a Fourier-transformed basis,
\begin{equation}
   \Psi^\dagger_k=\begin{pmatrix}
    f^\dagger_{A_k} && f^\dagger_{B_k}&& f_{A_{-k}} && f_{B_{-k}}
\end{pmatrix},
\nonumber
\end{equation}
leading to
\begin{equation}
H=\sum_k \Psi^\dagger_k \begin{pmatrix}
    0 && A_k^* && 0 && A_k^*\\
    A_k && 0 &&-A_k && 0\\
    0 && -A_k^* && 0 && -A_k^*\\
    A_k && 0 && -A_k &&0
\end{pmatrix}\Psi_k ,
\end{equation}
where
\begin{equation}
A_k=\frac{K_x}{8}+e^{ik}\frac{K_y}{8}. \nonumber
\end{equation}
The eigenvalues are given by 0 (doubly degenerate for every $k$) and $\pm \frac{\sqrt{K_x^2+K_y^2+2K_xK_y\cos(k)}}{4}$.

\begin{figure*}
 \includegraphics[width=2.3in]{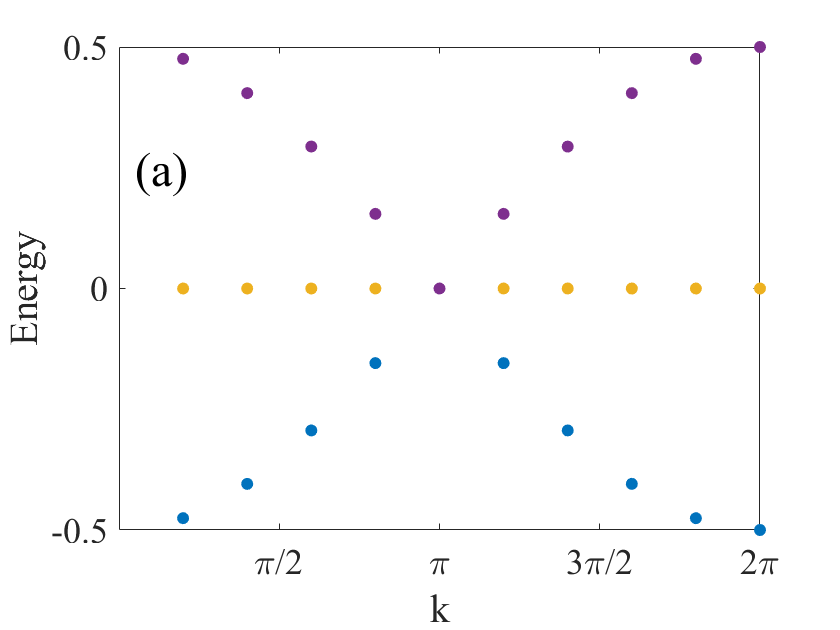}
 \includegraphics[width=2.3in]{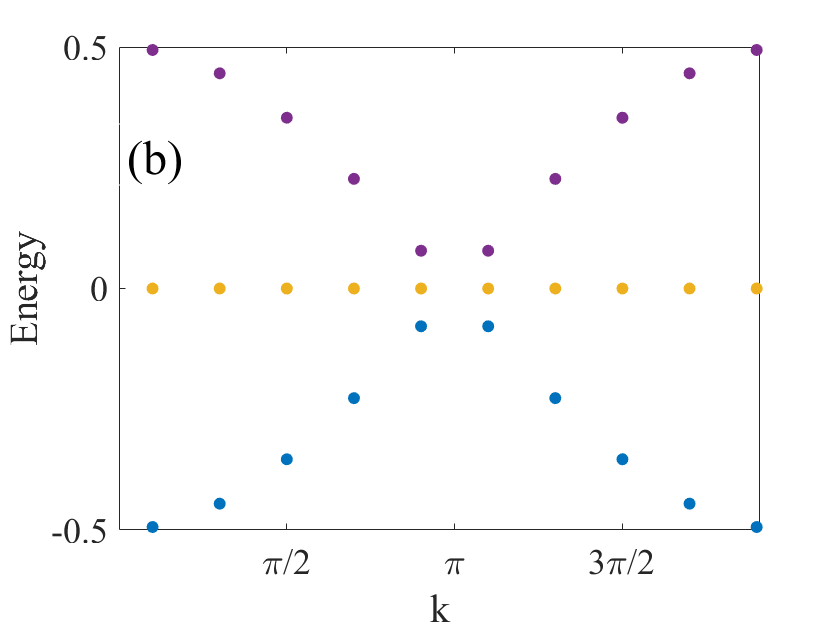}   
 \includegraphics[width=2.3in]{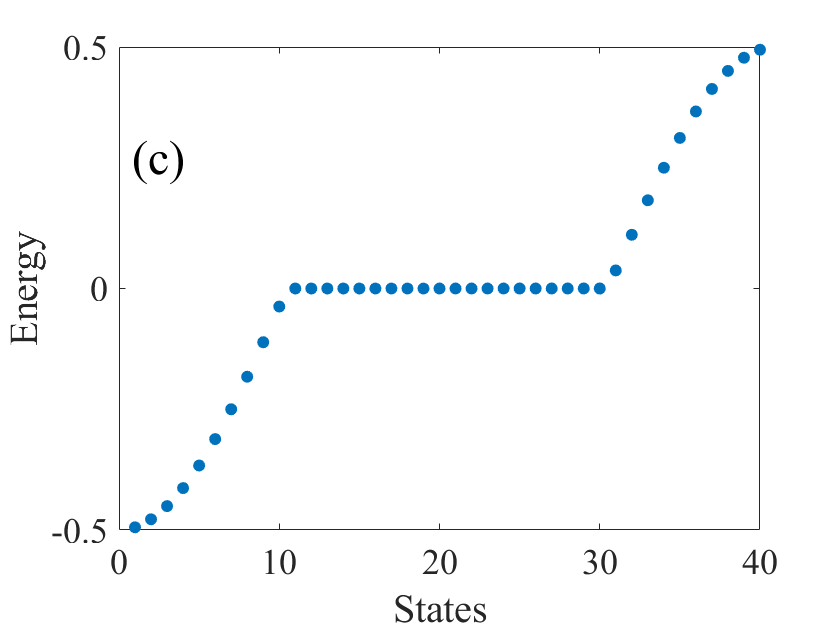}
 \caption{Jordan-Wigner spectra for $K_x=K_y=1$ and $N=20$. We show spectra for PBC in the odd-$\mathcal{N}$ sector (a), 
 PBC in the even-$\mathcal{N}$ sector (b)
 and for OBC (c). For PBC, energies are plotted against momentum $k$. The zero-energy modes here are doubly degenerate. The PBC ground state lies in the even-$\mathcal{N}$ sector.
 With OBC, as $k$ is not a good quantum number, we plot energies in increasing order. 
}
 \label{fig. JW iso}
\end{figure*}

\begin{figure*}
 \includegraphics[width=2.3in]{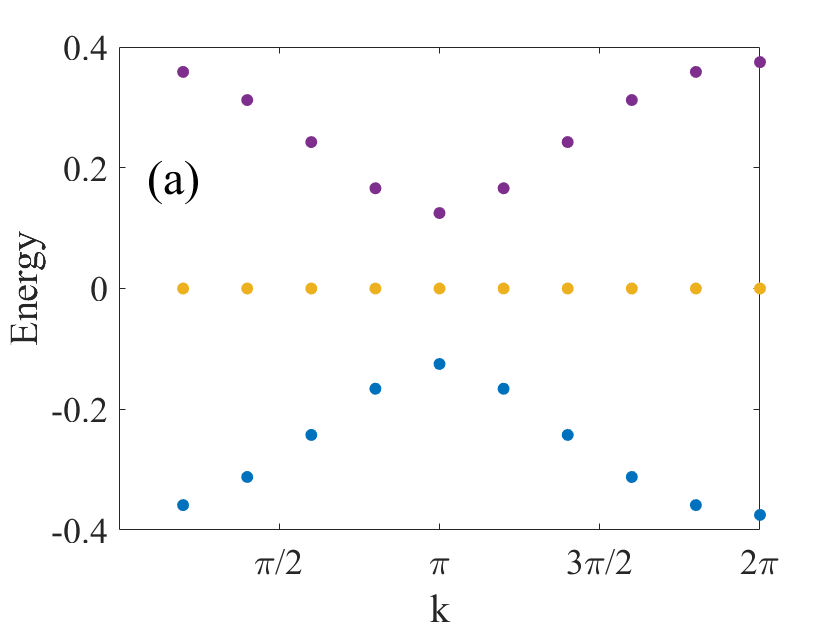}
 \includegraphics[width=2.3in]{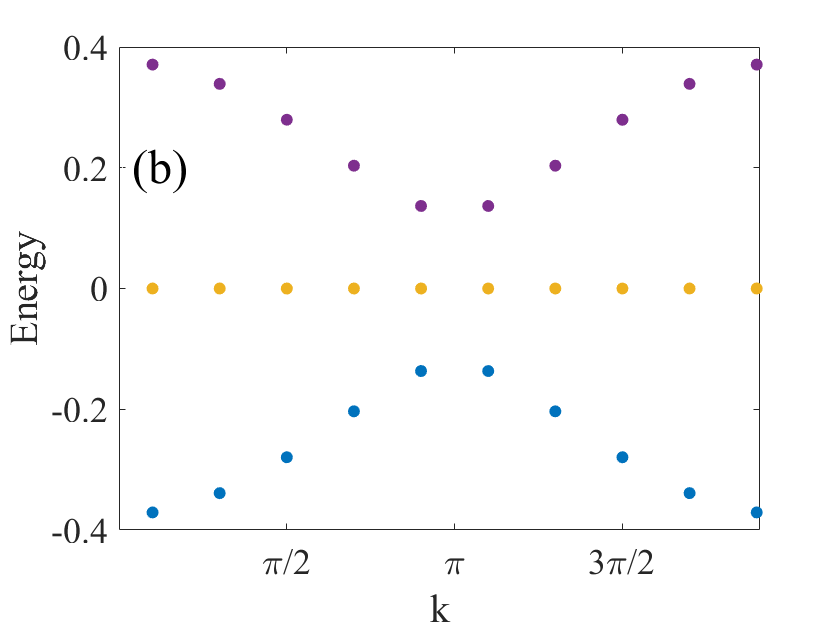}   
 \includegraphics[width=2.3in]{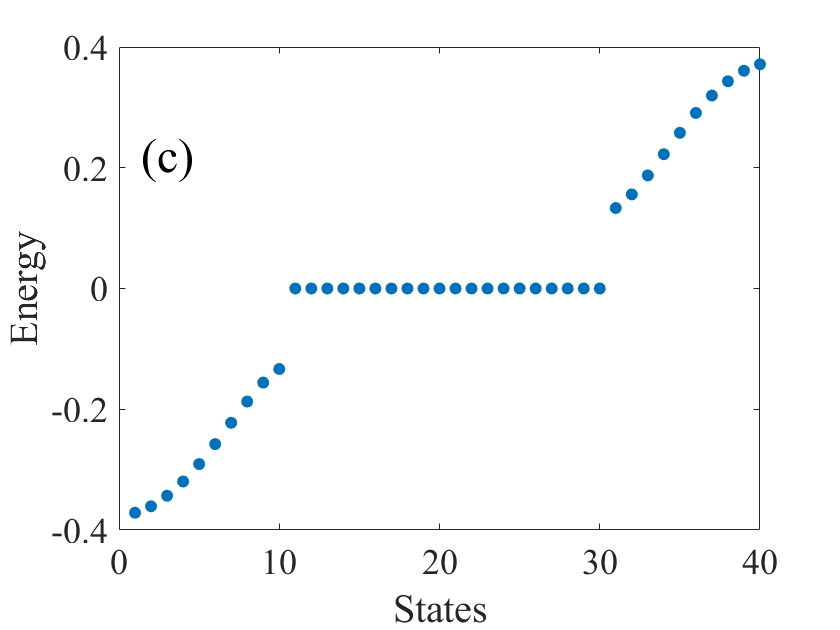}
 \caption{Jordan-Wigner spectra for $K_x=1$,$K_y=0.5$ and $N=20$ for PBC in the odd-$\mathcal{N}$ sector (a), 
 PBC in the even-$\mathcal{N}$ sector (b)
 and for OBC (c). With PBC, the ground state lies in the even-$\mathcal{N}$ sector. The zero modes here are doubly degenerate. Note that the dispersing bands are gapped. 
}
 \label{fig. JW aniso}
\end{figure*}

To determine the allowed values of $k$, we distinguish two cases: even or odd $\mathcal{N}$. These cases differ in the sign present in Eq.~\ref{eq.per}. 
For $\mathcal{N}$ odd, we have a translationally symmetric Hamiltonian with standard periodic  boundaries. This allows us to fix $k=\frac{4n\pi}{N}$. For $\mathcal{N}$ even, we have a negative sign in the couplings on a single bond. We assert that this can be treated as a translationally symmetric Hamiltonian, but with anti-periodic boundary conditions. This allows us to fix $k=\frac{2(2n-1)\pi}{N}$. 
For both even and odd cases, $n=1,2...,\frac{N}{2}$. The resulting Jordan-Wigner spectra are shown in Figs.~\ref{fig. JW iso} and ~\ref{fig. JW aniso}.

To find the ground state, we construct many-particle states by filling single-particle levels so as to produce the lowest total energy. We consider odd and even $\mathcal{N}$ sectors separately. In each sector, we choose $k$ values appropriately and fill an even/odd number of single-particle-levels. In both sectors, we find a large ground state degeneracy arising from zero modes. We compare ground state energies obtained from both sectors and select the lower one as the true ground state. We identify the ground state from the other sector as a low-lying excitation. 

We find that the true ground state lies in the even-(odd-)$\mathcal{N}$ sector when $\frac{N}{2}$ is even (odd). 
The ground state degeneracy is $2^{N/2-1}$ for all $N$ and all $K_y/K_x$. For $K_x\neq K_y$, the other sector yields a low-lying excited state with degeneracy $2^{N/2-1}$. 
In the thermodynamic $N\rightarrow\infty$ limit, the energy difference between the even- and odd-$\mathcal{N}$ sectors vanishes. This leads to a ground state degeneracy of $2^{N/2}$.

\begin{figure*}
 \includegraphics[width=3in]{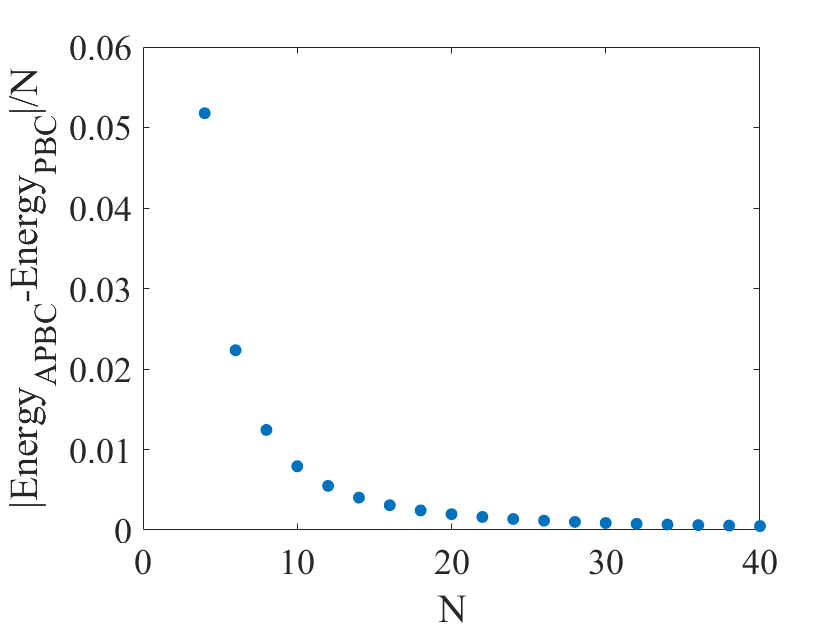}
\includegraphics[width=3in]{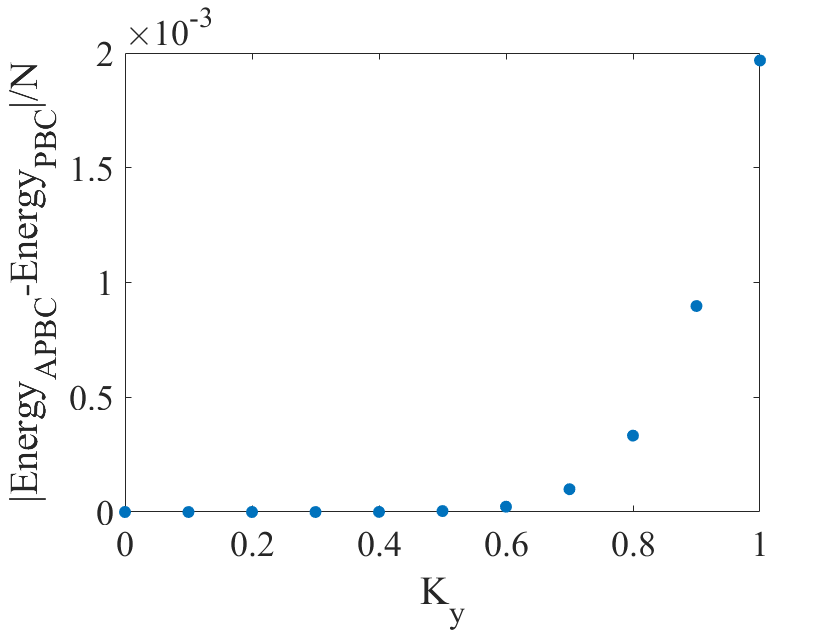}

    \caption{
    Difference in ground state energy between even- and odd-$\mathcal{N}$ sectors. Left: Energy difference as a function of system size for $K_x=K_y=1$. Right: Energy difference as a function of $K_y$, with $K_x$ set to unity at $N$ held fixed at 20. The plots show that the two sectors become degenerate as $N\rightarrow\infty$ or as $K_y\rightarrow0$}.   
\label{fig.JW_iso_vsN_vsK}
\end{figure*}

For $K_x=K_y$, the ground state sector has a degeneracy of $2^{N/2-1}$. The other sector has a higher degeneracy of $2^{N/2}$. This is due to the gapless nature of dispersive bands in this sector, see Fig.~\ref{fig. JW iso}(a), consistent with the results of Ref.~\onlinecite{Brzezicki2007}.
In the $N\rightarrow\infty$ limit, the ground states of the two sectors become degenerate as seen from Fig.~\ref{fig.JW_iso_vsN_vsK}(left). This yields a lower bound of $3\times 2^{N/2-1}$ for the ground state degeneracy as $N\rightarrow\infty$.  

\subsection{Summary of Jordan-Wigner results}
We summarize properties of the $S=1/2$ model, based on Jordan-Wigner fermionization. 
\begin{itemize}
\item With open boundaries, the ground state degeneracy is $2^{N/2}$.
\item With periodic boundaries, the ground state degeneracy is $2^{N/2-1}$.
\item In the thermodynamic limit, for $K_x\neq K_y$, the ground state degeneracy grows as $2^{N/2}$.
\end{itemize}

\section{Spin-basis wavefunctions}
\label{sec.spinbasis}

\begin{figure*}
    \includegraphics[width=0.45\linewidth]{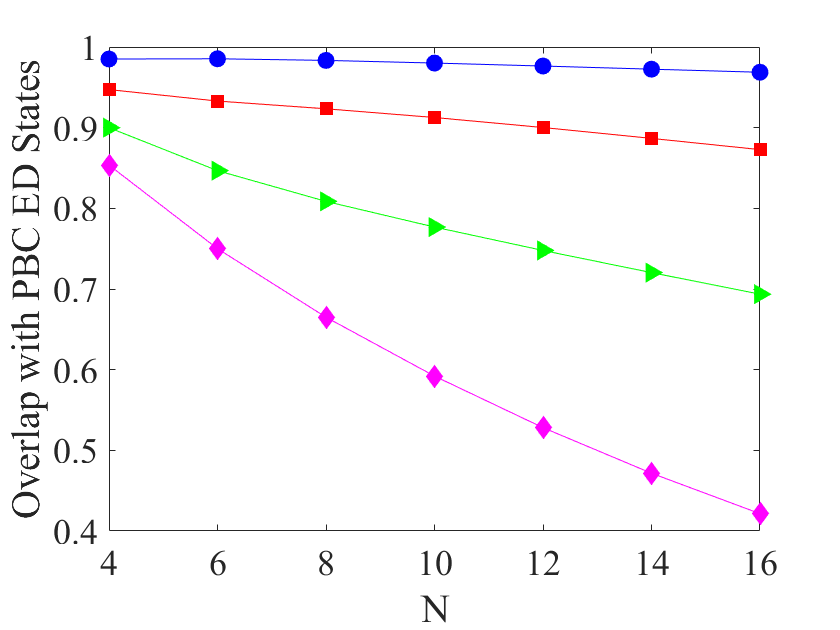}
    \includegraphics[width=0.45\linewidth]{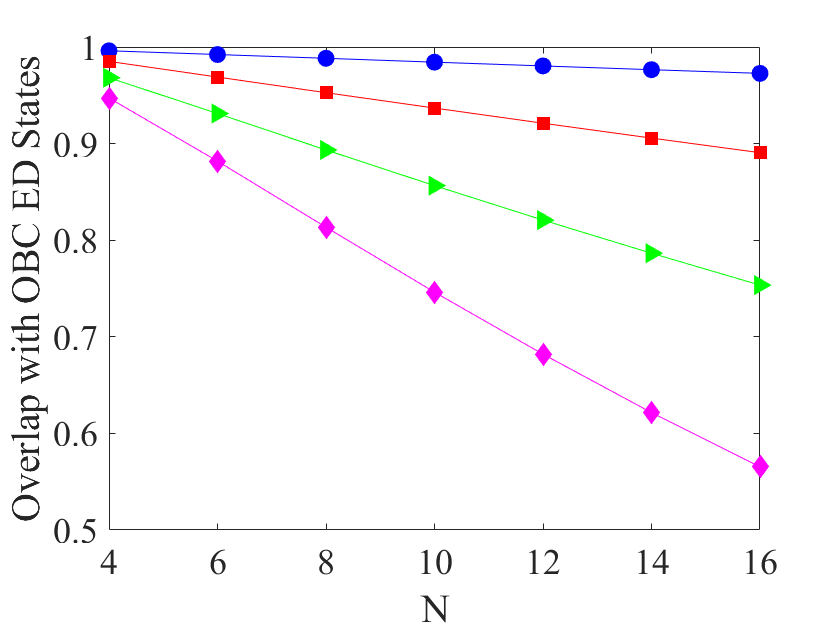}
    \caption{Overlap of the proposed wavefunctions with the ground state space obtained from exact diagonalization for PBC (left) and OBC (right). We fix $K_x=1$ and plot the overlap as a function of system size. 
    We show data for $K_y=0.25$ (blue), 0.5 (red), 0.75 (green) and 1 (magenta). 
    The quantity on the Y-axis corresponds to $\sum_i \vert \langle \psi \vert \mathrm{ED}_i\rangle \vert^2$, where $\vert  \psi \rangle$ is any one of the proposed wavefunctions and $\vert \mathrm{ED}_i\rangle$ are the ground states obtained from exact diagonalization.}
    \label{fig. Overlap spin half basis states}
\end{figure*}

 \begin{figure*}
    
    \includegraphics[width=0.45\linewidth]{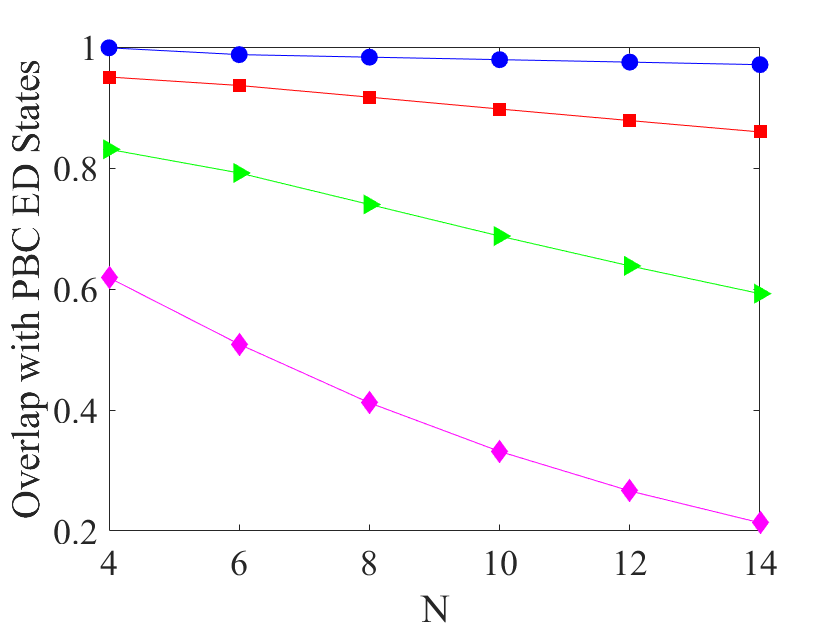}
    \includegraphics[width=0.45\linewidth]{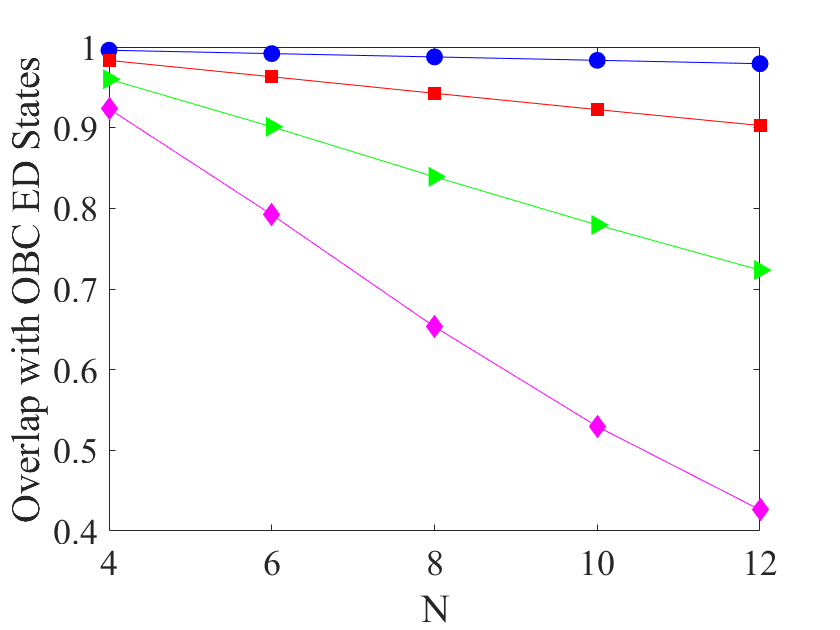}
    \caption{Overlap of the proposed wavefunction(s) with that obtained from exact diagonalization for $S=1$ with PBC (left) and OBC (right). We fix $K_x=1$ and plot the overlap as a function of system size. We show data for $K_y=0.25$ (blue), $K_y=0.5$ (red), $K_y=0.75$ (green) and $K_y=1$ (magenta).
    For PBC, the overlap is computed for a unique ground state with $\vert t_{S,X}\rangle$ on every $X$ bond.   
For OBC, the overlap is evaluated for four states that are constructed as follows. The bonds in the bulk are placed in $\vert t_{S,X}\rangle$. The two edge bonds could be in either $\vert t_{S,X}\rangle$ or $\vert s_{S,X}\rangle$ states. Each of the four resulting states yields the same overlap with the ground states of exact diagonalization. 
}
    \label{fig. Overlap spin one basis states}
\end{figure*}
We now present analytic forms for the ground state wavefunctions in the spin basis. We first present our results for $S=1/2$, with extensions to general $S$ values discussed subsequently. 
As building blocks for our $S=1/2$ wavefunctions, we define four bond-states:
\begin{itemize}
    \item $\vert s \rangle = \frac{1}{\sqrt{2}}\{\vert \!\uparrow,\downarrow\rangle -\vert \!\downarrow,\uparrow\rangle\} $: the two sites at the ends of a bond form a spin-singlet. To write this wavefunction, we must fix a certain order of the sites. This choice does not alter our final result, except for an overall negative sign.
    \item $\vert t_x \rangle = \frac{1}{\sqrt{2}}\{\vert\! \uparrow,\uparrow\rangle -\vert \!\downarrow,\downarrow\rangle\} $: We call this a triplet-x state. It is defined so as to produce a zero net moment along the x direction.
    \item $\vert t_y \rangle = \frac{i}{\sqrt{2}}\{\vert \!\uparrow,\uparrow\rangle +\vert\! \downarrow,\downarrow\rangle\} $: This triplet-y state produces a zero net moment along the y direction.
    \item $\vert t_z \rangle = \frac{1}{\sqrt{2}}\{\vert\!\uparrow,\downarrow\rangle +\vert \!\downarrow,\uparrow\rangle\} $: This is the usual triplet-z state, with zero net moment along the z direction.
\end{itemize}
To see the utility of these states, we first consider an isolated X bond, with two spins coupled by an XX Ising interaction. The singlet and triplet states defined above serve as eigenstates of the bond-Hamiltonian. The ground state is doubly degenerate, consisting of $\vert s \rangle$ and $\vert t_x \rangle$. The first excited state is also doubly degenerate, consisting of $\vert t_y \rangle$ and $\vert t_z \rangle$. 
Below, in Sec.~\ref{sec.pert}, we will use this single-bond spectrum to derive the ground state for the Kitaev chain using a perturbative scheme. 

We next present wavefunctions for the chain, assuming $K_x > K_y$. We propose bond-factorized states, where each X bond is assigned a singlet or triplet-x state. We then construct ground states as follows. 
\begin{itemize}
\item Open boundaries: For concreteness, we assume that both end-bonds are of the X type. On each X bond, we assign a singlet or a triplet-x. We have a twofold choice on each bond, leading to $2^{N/2}$ wavefunctions. Each is a ground state of the spin chain. 
\item Periodic boundaries: Any `even-triplet' state is a ground state. These are bond-factorized states where an even number of X-bonds are in the triplet-x state, while the rest are in the singlet state. For example, for $N=4$, we may have zero triplets or two triplets. The constraint of an even number of triplets leads to $2^{N/2-1}$ ground states --- see App.~\ref{app.even}.

\end{itemize}

We emphasize that these states are exact ground states only when $K_y\rightarrow 0$. In general, they are approximate forms that capture the dominant weight of the true ground states. Fig.~\ref{fig. Overlap spin half basis states} shows the overlap of these states with the ground state space obtained from exact diagonalisation. We obtain overlaps for system sizes from $N=4$ to $N=16$ for various values of $K_y/K_x$. The overlap is stronger for smaller system sizes and higher anisotropies.  We justify this observation using a perturbative approach in Sec.~\ref{sec.pert} below.

\subsection{Wavefunctions for half-integer spin-$S$}
\label{sec.halfintS}

For $S>1/2$, we define bond wavefunctions on the X bonds on the same lines as for $S=1/2$. It can be easily seen that even for arbitrary $S$, an isolated X bond has two degenerate ground states. We take them to be analogues of the singlet and triplet-x states, defined as
\begin{itemize}
\item $\vert s_{S,X} \rangle = \frac{1}{\sqrt{2}}\{\vert S,-S\rangle -\vert -S,S\rangle\}$, an `anti-symmetric' state, and
\item $\vert t_{S,X} \rangle = \frac{1}{\sqrt{2}}\{\vert S,-S\rangle +\vert -S,S\rangle\}$, a `symmetric' state. 
  \end{itemize}
Here, $\vert \pm S \rangle$ are single-spin states that are polarized along $\pm$X. Note that $\vert t_{S,X} \rangle$ and $\vert s_{S,X}\rangle$ 
reduce to the usual singlet and triplet-x, when $S=\frac{1}{2}$.

The ground states for the Kitaev chain are defined in direct analogy with the $S=1/2$ case. We construct spin-$S$ bond-factorized wavefunctions as follows. On each X bond, 
in place of the singlet state for $S=1/2$, we use $\vert s_{S,X} \rangle$. In place of the triplet-x state for $S=1/2$, we use $\vert t_{S,X}\rangle$. With open boundaries, we have $2^{N/2}$ ground states as each X bond can either be in $\vert s_{S,X} \rangle$ or $\vert t_{S,X}\rangle$ states. With periodic boundaries, we have 2$^{N/2-1}$ ground states, defined as bond-factorized wavefunctions with an even number of $\vert t_{S,X} \rangle$'s (in analogy with the even-triplet rule for $S=1/2$).  

These states can be characterized in terms of $W$'s, the conserved quantities defined in Eq.~\ref{eq.Ws} above. With half-integer $S$, conserved quantities ($W$'s) are defined only on the X bonds. 
The bond states $\vert s_{S,X}\rangle$ and $\vert t_{S,X}\rangle$ correspond to $W=+1$ and $-1$ respectively.
With open boundaries, we may independently assign $W$ to be $+1$ or $-1$ on each X-bond. With periodic boundaries, we impose a constraint that the product of all $W$'s must be $+1$, equivalent to having an even number of triplets.

\subsection{Wavefunctions for integer spin-$S$}
\label{sec.intS}

With periodic boundaries, we propose a unique ground state that is expressed as a bond-factorized wavefunction. On each X bond, we place the symmetric state $\vert t_{S,X} \rangle$ when $K_x>K_y$. 
The isotropic $K_x=K_y$ limit is discussed separately in Sec.~\ref{sec.isotropic} below. With integer-$S$, conserved quantities ($W$'s) are defined on the X bonds as well as the Y bonds. On the X bonds, $\vert t_{S,X} \rangle$ corresponds to $W=+1$. Our proposed ground state is left unchanged by the action of the $W$'s on Y bonds as well. As a result, $W$'s are uniformly $+1$ in the proposed ground state. With open boundaries, the bonds in the bulk are placed in $\vert t_{S,X}\rangle$. The two edge bonds could be in either $\vert t_{S,X}\rangle$ or $\vert s_{S,X}\rangle$ states, giving rise to four ground states.

In Fig.~\ref{fig. Overlap spin one basis states}, we take the simplest example of $S=1$. We plot the overlap of the proposed ground state with that obtained from exact diagonalization. We show results for system sizes $N=4$ to $N=14$ and 4 values of $K_y/K_x$. As with the case of $S=1/2$, the overlap decreases with system size and increases with anisotropy -- in line with perturbative discussion of Sec. ~\ref{sec.pert}.

\section{Perturbation theory}
\label{sec.pert}
We now rationalize the wavefunction forms that were presented in Sec.~\ref{sec.spinbasis} using a perturbative approach. Here, a rigorous Schrieffer-Wolff-like presentation is cumbersome due to large ground state degeneracy in the case of half-integer spins. We instead describe the structure of the perturbative expansion, focusing on processes that determine the ground state. We consider the anisotropic limit where $K_x \gg K_y$. We will later argue that our conclusions hold as long as $K_x > K_y$. The discussion below can be easily adapted to the case where $K_y > K_x$, with the $K_x=K_y$ case discussed separately in Sec.~\ref{sec.isotropic}. The results can also be extended to regimes where $K_x$ and/or $K_y$ are negative by performing local spin rotations.

We first address the $S=1/2$ problem with periodic boundaries. 
Starting from the $K_y \rightarrow 0$ limit, we consider an isolated $X$ bond. The bond Hilbert space is four-dimensional. With the bond Hamiltonian given by $S_{A_i}^xS_{B_i}^x$, the Hilbert space splits into two doublets: the singlet and triplet-x are the twofold ground states, while triplet-y and triplet-z are the two degenerate excited states. The energy separation between the doublets is $K_x/2$.  

With $K_y \rightarrow 0$, the Kitaev chain has $\frac{N}{2}$ decoupled bonds, each of the $X$ type. The energy on each bond can be independently minimized by placing it in the singlet or triplet-x state. This leads to a ground state degeneracy of $2^\frac{N}{2}$. We next turn on the $K_y$ coupling as a perturbation, following the approach of Ref.~\cite{gordon2022insights}.

Each $K_y$ term acts on two X bonds that are separated by a Y bond. The perturbing operator is given by $S_{B_i}^y S_{A_{i+1}}^y$, following the labeling convention in Eq.~\ref{eq.Hamiltonian}. Its action on the bond-states is given by
\begin{eqnarray}
    S_{B_i}^yS_{A_{i+1}}^y |s_0\rangle_{A_iB_i}\rightarrow -\frac{1}{4}|t_y\rangle_{A_iB_i}, \nonumber 
    \\
    S_{B_i}^yS_{A_{i+1}}^y |t_x\rangle_{A_iB_i}\rightarrow i\frac{1}{4}|t_z\rangle_{A_iB_i}, \nonumber \\S_{B_{i-1}}^yS_{A_{i}}^y |s_0\rangle_{A_iB_i}\rightarrow \frac{1}{4}|t_y\rangle_{A_iB_i}, \nonumber \\
    S_{B_{i-1}}^yS_{A_{i}}^y |t_x\rangle_{A_iB_i}\rightarrow i\frac{1}{4}|t_z\rangle_{A_iB_i}, \nonumber \\
    S_{B_i}^yS_{A_{i+1}}^y \vert t_y\rangle_{A_iB_i}\rightarrow -\frac{1}{4}\vert s_0\rangle_{A_iB_i}, \nonumber \\
    S_{B_{i-1}}^yS_{A_{i}}^y \vert t_y\rangle_{A_iB_i}\rightarrow \frac{1}{4}\vert s_0\rangle_{A_iB_i}, \nonumber \\
    S_{B_i}^yS_{A_{i+1}}^y\vert t_z\rangle_{A_iB_i} \rightarrow i\frac{1}{4}\vert t_x\rangle_{A_iB_i}, \nonumber \\
    S_{B_{i-1}}^yS_{A_{i}}^y \vert t_z\rangle_{A_iB_i} \rightarrow i\frac{1}{4} \vert t_x\rangle_{A_iB_i}.
\end{eqnarray}
These expressions lead to a highly structured perturbative expansion. We have contributions from two types of processes: local and global. The former has been discussed in Ref.~\onlinecite{gordon2022insights}, but not the latter. 

\begin{figure*}
\includegraphics[width=3in]{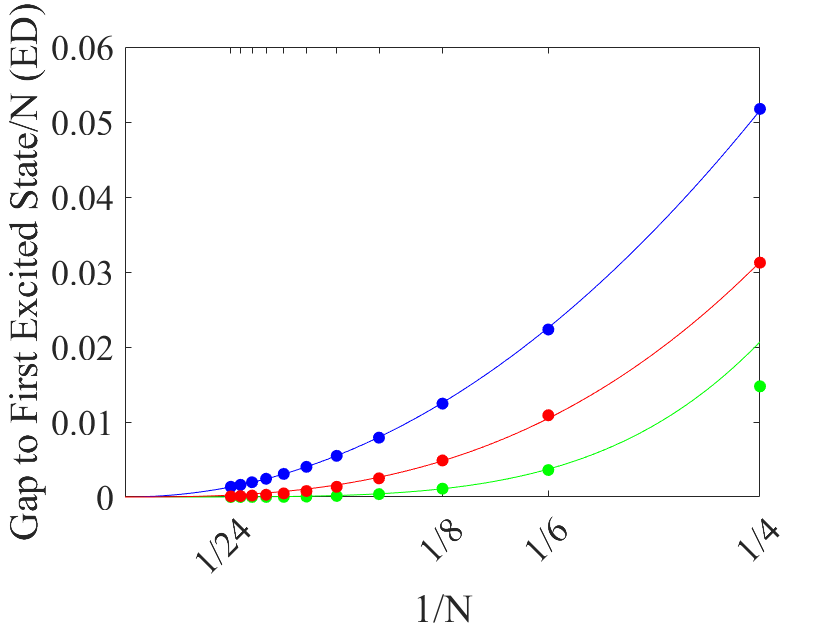}
\includegraphics[width=3in]{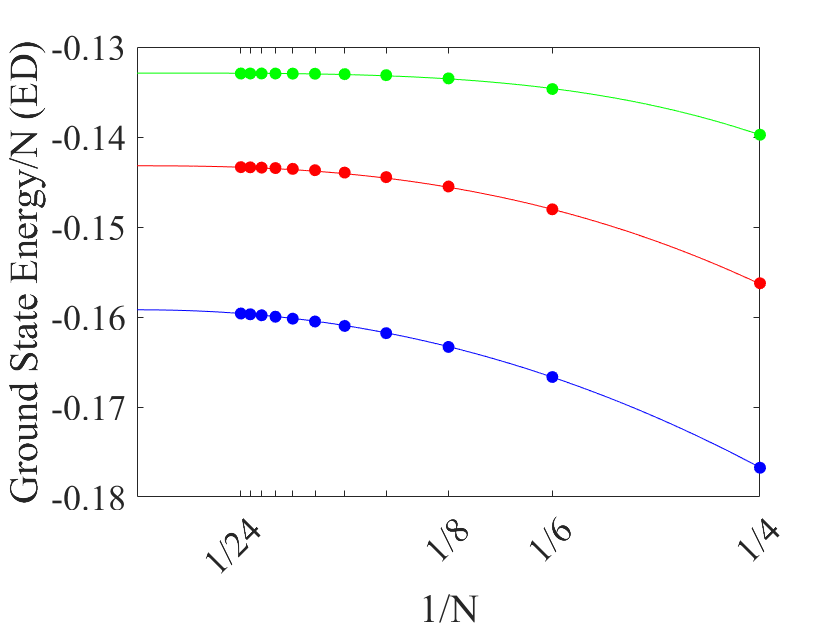}

\caption{Gap (left) and ground state energy (right) obtained from exact diagonalization for the $S=\frac{1}{2}$ chain with PBC. Energies are presented on a per-site basis. We fix $K_x=1$ and present data for $K_y=0.5$ (green), $K_y=0.75$ (red) and $K_y=1$ (blue). The plots show power-law fits that allow for extrapolation to the $N\rightarrow\infty$ limit. The gap goes to zero in the thermodynamic limit. Ground state energy (per site) converges to -0.1329 for $K_y=0.5$, -0.1432 for $K_y=0.75$ and -0.1592 for $K_y=1$. These are in good agreement with the perturbative result in Eq.~\ref{eq.leading}.}
\label{fig. Gap and GS energy PBC spin half}
\end{figure*}

\begin{figure}
\includegraphics[width=\columnwidth]{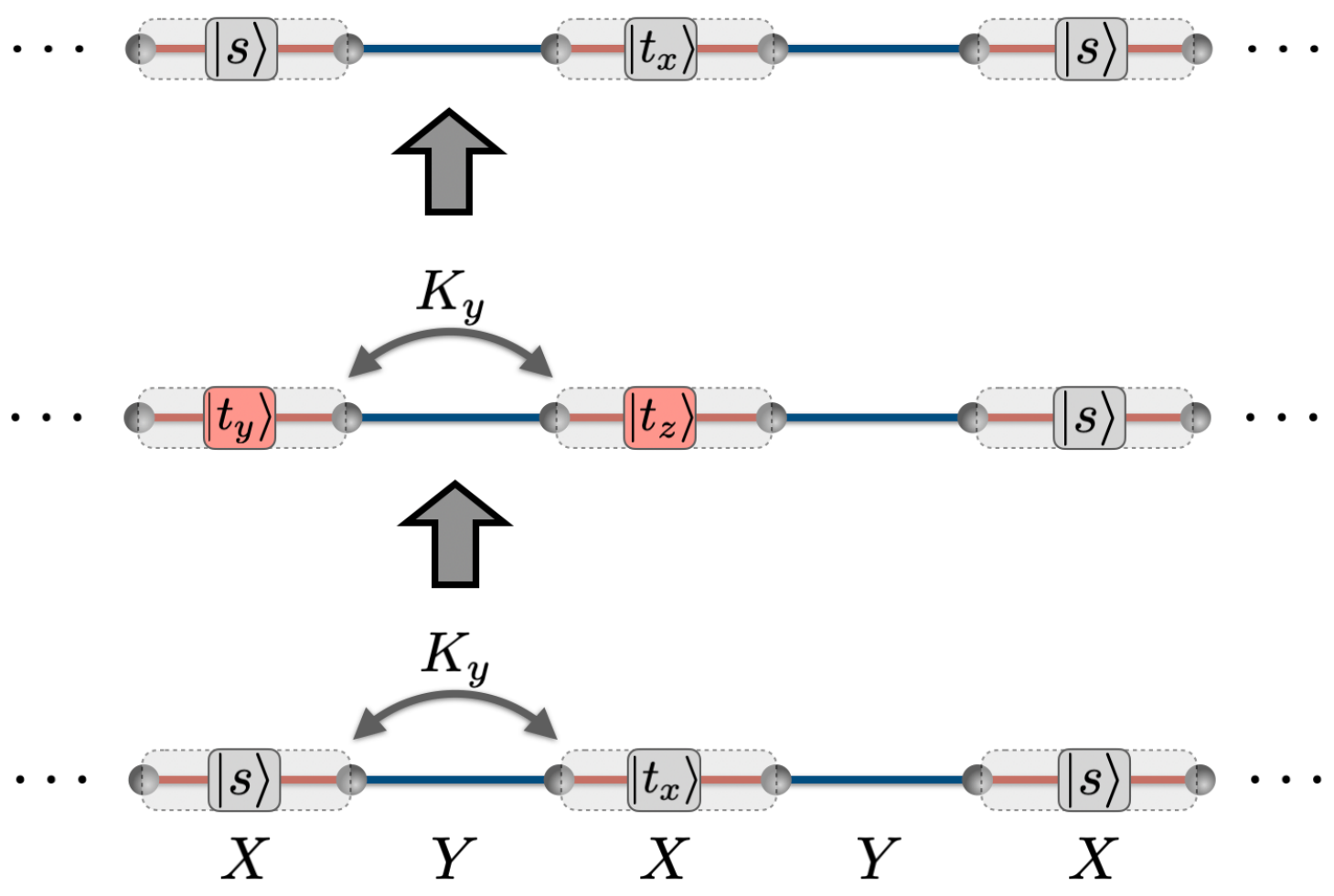}
\caption{The local process for $S=1/2$ in the perturbative expansion. We begin with a low-energy state, where every X bond is either in the $\vert s\rangle$ or the $\vert t_x \rangle$ state. The $K_y$ coupling on a Y bond excites two adjacent X bonds. With two successive applications, we return to initial low-energy state. }
\label{fig.locprocess}
\end{figure}

Local processes arise at the second order. With a single action of the perturbing operator, two X-bonds are excited from their respective bond-ground states to bond-excited states. This produces a virtual state with an energy cost of $K_x$. To return to the ground state, we act the same $K_y$ term once again. An example of a local process is shown pictorially in Fig.~\ref{fig.locprocess}. Crucially, the second order processes act in the same way for any initial ground state of the Kitaev chain. That is, all $2^{N/2}$ ground states of the $K_y \rightarrow 0$ limit acquire the same energy correction at second order. 

The local process provides the leading-order correction to the ground state energy. 
Fig. \ref{fig. Gap and GS energy PBC spin half} shows the ground state energy in the thermodynamic limit, obtained by extrapolating exact diagonalization results to $N\rightarrow \infty$. 
We compare this with the leading-order perturbative result,
\begin{equation}
    \frac{E}{N}\sim \frac{-K_x}{8}-\frac{ K_y^2}{32K_x}.
    \label{eq.leading}
\end{equation}
Here, the zeroth order energy $(-K_x/8)$ is obtained by minimizing the energy on each $X$ bond. The leading correction scales as $K_y^2$, with an energy denominator of $K_x$ -- the cost of two `defect' bonds generated by the local process. The correction is multiplied by $N/2$, the number of $K_y$ bonds.   
To compare with exact diagonalization results, we set $K_x=1$. With $K_y=0.5$, we obtain $\frac{E}{N}\sim-0.1328$; with $K_y=0.75$, we obtain $\frac{E}{N}\sim -0.1426$ and with $K_y=1$, we have $\frac{E}{N}\sim -0.1562$. These estimates are within 2\% of the $N\rightarrow\infty$ values quoted in Fig. \ref{fig. Gap and GS energy PBC spin half}.

\begin{figure}
\includegraphics[width=\columnwidth]{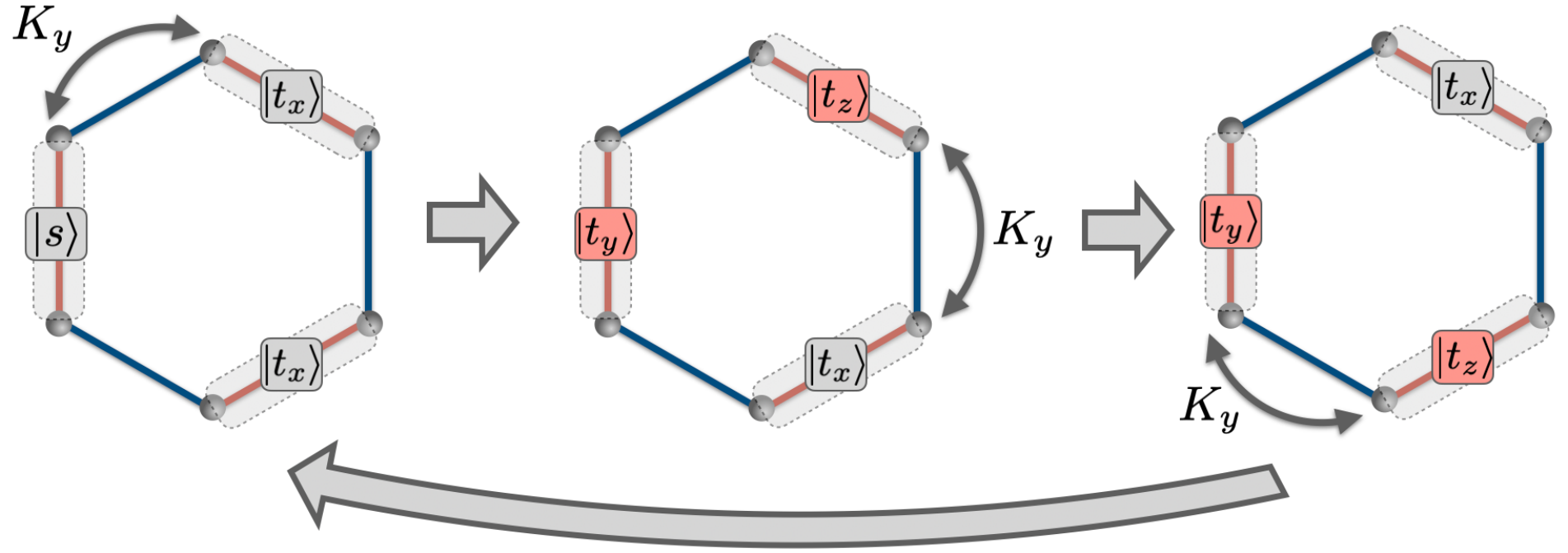}
\caption{The global process for $S=1/2$ in the perturbative expansion, depicted for N=6. Each $K_y$ coupling flips two adjacent X bonds -- from a ground state to an excited state or the other way around. We return to the initial low-energy state after the $K_y$ coupling acts on every Y bond in succession. }
\label{fig.globprocess}
\end{figure}

We next consider the `global' process as shown pictorially in Fig.~\ref{fig.globprocess}. This term occurs at $(N/2)^{\mathrm{th}}$ order in $K_y$. The $K_y$ term acts once on every Y bond as we go around the chain (in any order). At the end of this process, each X-bond comes back to precisely the same bond-ground state that it started with. However, the wavefunction may pick up an overall sign. To see this, we note that $\vert s_0\rangle_{A_iB_i}$ picks up a negative sign when acted by perturbations at both ends of the bond. However, the triplet-x does not pick up a sign. As a result, the overall wavefunction picks up a negative sign if we have an odd number of singlets. 
At the same time, each step in the process introduces an energy denominator that is negative. The overall energy denominator has a sign given $(-1)^{\frac{N}{2}-1}$.

On the basis of these observations, we obtain the following result. The global process splits the ground states of the $K_y \rightarrow 0$ limit into two degenerate sets, each with $2^{N/2-1}$ elements. One set contains all low-energy states with an even number of bond-triplets. These states have an even (odd) number of singlets if $N/2$ is even (odd). Considering the sign of the wavefunction and the energy denominator, these states enjoy a net lowering of energy due to the global process. The second set contains states with an odd number of bond-triplets; these states suffer an energy increase.

We make three key observations about the nature of the perturbation theory: (i) Although the ground state is highly degenerate in the $K_y\rightarrow 0$ limit, we may employ non-degenerate perturbation theory. This is because perturbations do not couple distinct ground states at any order. This simplification arises from our choice of expressing ground states as $|s_{S,X}\rangle$ and $|t_{S,X}\rangle$ arrangements.  (ii) A perturbative process at any order can be expressed as a combination of local processes and the global process. (iii) As the qualitative arguments (local processes do not distinguish among ground states, global process splits ground states into two degenerate sets) above hold at all orders in perturbation theory, we conjecture that our conclusions are valid for any value of $K_y/K_x$, as long as $K_x > K_y$ to ensure that the initial states are ground states of the unperturbed $K_y\rightarrow 0$ Hamiltonian. These conclusions are supported by our numerical ED results as well as the exact Jordan-Wigner solution for $S=\frac{1}{2}$.

On the basis of these arguments, we find that the ground state degeneracy is reduced from $2^{N/2}$ in the unperturbed limit to $2^{N/2-1}$.
The ground states are constructed by placing valence bonds (singlet or triplet-x) on X bonds, subject to a global constraint: the total number of triplets must be even. The lowest excited states are constructed in similar fashion, but with an odd number of triplets. As the excitation gap to these latter states is a perturbative effect at order $(N/2)$, it vanishes when $N\rightarrow \infty$. We arrive at a $2^{N/2}$-fold ground state degeneracy in the thermodynamic limit.

The case of open boundaries follows immediately. We consider the dangling bonds at the ends to be of the X type for concreteness. 
With open boundaries, local processes as shown in Fig.~\ref{fig.locprocess} can play a role. However, the global process does not apply. It is not possible to return to the ground state by acting the $K_y$ term on every Y bond -- as we require perturbations to act on either end of each X bond. Therefore, all perturbative processes are merely combinations of the local second-order process. As this process does not differentiate among low-energy states, we conjecture that all $2^{N/2}$ ground states remain degenerate to all orders in perturbation theory.

\subsection{Extension to half-integer spin-$S$}
We begin with the $K_x\gg K_y$ limit, where we have isolated X bonds. Each X bond has two ground states given by $\vert s_{S,X}\rangle$ and $\vert t_{S,X} \rangle$ as defined in Sec.~\ref{sec.spinbasis} above. The perturbative $K_y$ term acts on a Y bond, taking the two adjacent X bonds to excited states. In analogy with the $S=1/2$ case, we have local and global perturbative processes. 

The lowest-order correction appears at second order, a local process analogous to the $S=\frac{1}{2}$ case where the same perturbing term is applied twice to return to the ground state manifold.

For $S=1/2$ and periodic boundaries, the global process appears at order $(N/2)$. It can be viewed as going around the loop (generated by periodic boundaries), applying the $K_y$ term on each Y bond. For $S>1/2$, an analogous global process appears at order $(N/2)(2S)$. It requires going around the loop $2S$ times, so that we operate on each Y bond $2S$ times. At the end of this process, each site that started out in the $\vert S \rangle$ state (fully polarized along X) reaches the $\vert -S \rangle$ state (polarized along -X) and vice versa. This brings us back to the initial ground state that we started with. However, the wavefunction may pick up an overall sign. This can be understood as follows. Each $\vert s_{S,X}\rangle$ bond-state picks up a negative sign at the end of the global process, while each $\vert t_{S,X} \rangle$ bond-state is unchanged. In the energy denominator associated with the global process, each step yields a negative sign. We obtain an overall sign given by $(-1)^{NS-1}$.

Based on these observations, we draw the following conclusions for the case of periodic boundaries. The global process splits the $2^{N/2}$-fold ground state into two sets, each with $2^{N/2-1}$ elements. States with an even number of triplets (i.e., $\vert t_{S,X} \rangle$ states) have their energy lowered, while states with an odd number of triplets show an energy increase. We obtain a ground state degeneracy of $2^{N/2-1}$. The lowest excited states have a gap, due to a perturbative effect at order $(N/2)(2S)$. In the thermodynamic $N\rightarrow \infty$ limit, the gap vanishes to yield a degeneracy of $2^{N/2}$. 

With open boundaries, the global process becomes inoperative. As a result, all $2^{N/2}$ states form the ground state manifold.

\subsection{Extension to integer spin-$S$}

Starting from isolated X bonds, we have local and global perturbative processes. Unlike the case of half-integer $S$, local processes immediately break the ground state degeneracy. 

\begin{figure}
\includegraphics[width=\columnwidth]{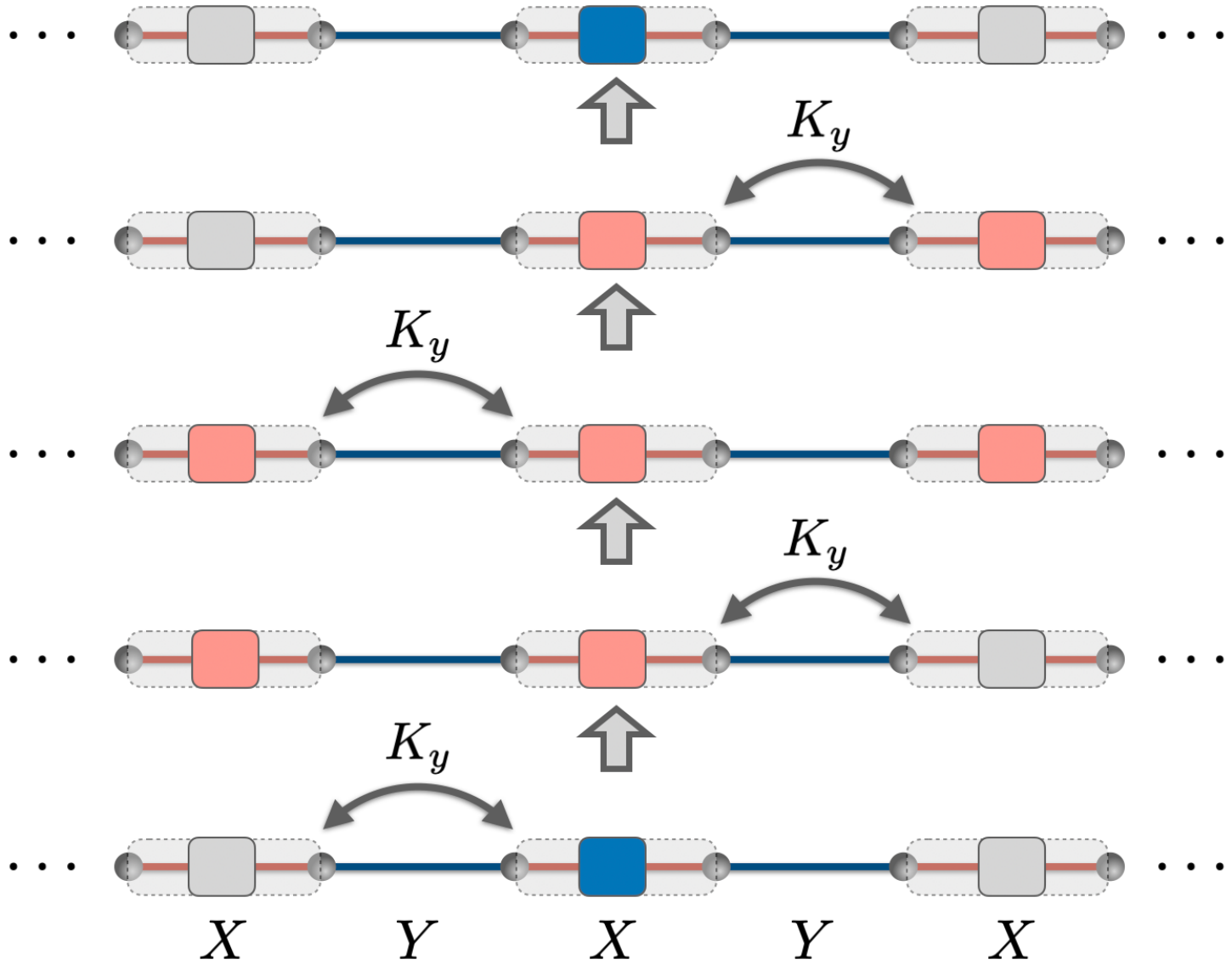}
    \caption{The lowest non-trivial correction for $S=1$. In the initial state, each X bonds is placed in one of two low-energy states. The central bond is shown in blue. Perturbations produce excited bond-states, shown in pink. At fourth order, we may return to a low-energy state. Crucially, this process contributes only if the central bond is initially in a $\vert t_{S,X}\rangle$ state. 
    }
    \label{fig.intSlocprocess}
\end{figure}

To illustrate this, we take $S=1$ as an example. We begin with a low-energy state where every X bond is assigned either a $\vert s_{S,X}\rangle$ or a $\vert t_{S,X}\rangle$ state. We treat the $K_y$ couplings as perturbations. Lowest order processes appear at second order in $K_y$, similar to the case shown in Fig.~\ref{fig.locprocess}. However, these processes provide a trivial contribution that is identical for all low-energy states. The lowest non-trivial process appears at fourth order -- shown in Fig.~\ref{fig.intSlocprocess}. It lowers the energy if we start from a $\vert t_{S,X}\rangle$ state. With a $\vert s_{S,X}\rangle$ state, the amplitude of this process vanishes due to destructive interference. This can be traced to the negative sign in the definition of $\vert s_{S,X}\rangle$ which leads to cancellation. For arbitrary spin-$S$, an analogue of this process appears at order $4S$.

Unlike the case of half-integer $S$, local processes favour $\vert t_{S,X}\rangle$ over $\vert s_{S,X}\rangle$. We conclude that energy is minimized with $\vert t_{S,X}\rangle$ on every bond, providing a unique ground state. With local processes playing a strong role, we may disregard any global processes as they occur at higher order.

With open boundaries, the local perturbative process of Fig.~\ref{fig.intSlocprocess} is effective in the bulk of the chain. It is not effective at the edge, as two Y bonds are required on either side of the reference X bond. Each dangling bond can then be placed in either $\vert t_{S,X}\rangle$ or $\vert s_{S,X}\rangle$. We are left with a fourfold-degenerate set of ground states.

\section{The isotropic limit}
\label{sec.isotropic}
The perturbative approach described above carries through for any $(K_x,K_y)$ as long as $ K_x   >  K_y $. The same approach can be easily adapted for $ K_y  > K_x $ by placing valence-bond-states on the Y bonds rather than X bonds. The isotropic point where $K_x=K_y$ poses an interesting challenge. At this point, the starting point of our perturbation theory -- that of minimizing energy on one set of bonds first -- no longer applies. Below, we restrict our attention to periodic boundaries for simplicity.  

\subsection{Half-integer $S$}
\begin{figure}
    \includegraphics[width=1\linewidth]{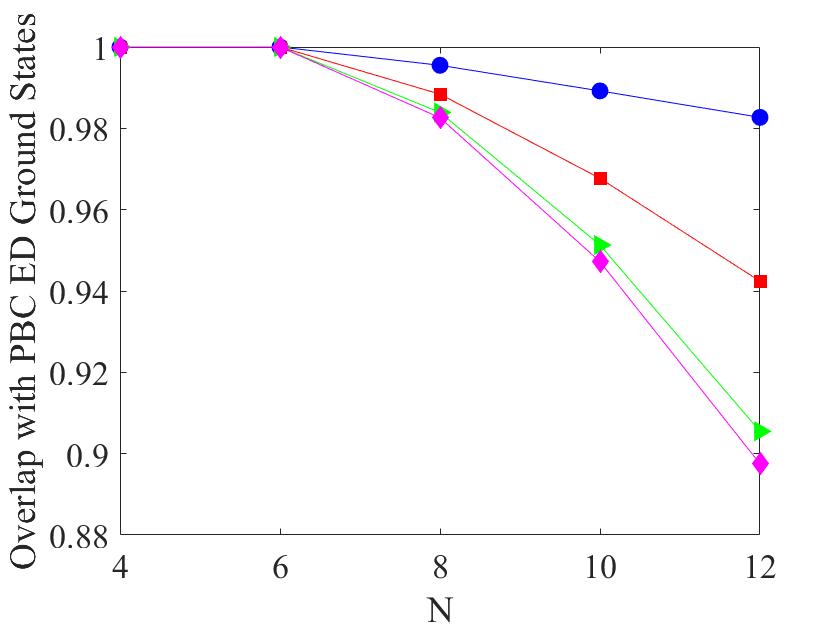}
  
    \caption{Overlap of ground states proposed for $K_x=K_y$, $S=1/2$ and PBC, with the ground state space obtained from exact diagonalization. 
    Overlaps are calculated for states obtained by solving a generalized eigenvalue problem. 
    Data for $K_x=K_y=1$ are shown in magneta. We also show data for anisotropic cases, where the same generalized-eigenvalue-problem approach is used. We show data for $K_y/K_x=0.25$ (blue), $K_y/K_x=0.5$ (red) and $K_y/K_x=0.75$ (green). For $K_x\neq K_y$, this approach increases the overlap compared to using a single valence bond arrangement.  }
    \label{fig. Overlap spin half reduced basis}
\end{figure}

We take the following approach for half-integer values of $S$. We consider two sets of low-energy states: those that minimize energy on X bonds and those that minimize Y bonds. Each set contains $2^{N/2}$ states. Crucially, states within each set are mutually orthogonal. 
However, a state from one set may have a non-zero overlap with any state from the other set. In order to determine the low-energy spectrum, we use a variational approach using valence bond states on X bonds as well as Y bonds. 

We denote the manifold of low-energy states as 
\begin{eqnarray}
\{\zeta\} = \Big\{\{x^{e}\},\{x^{o}\},\{y^{e}\},\{y^{o}\}\Big\},
\end{eqnarray}

$\{x^{e}\}$ and $\{x^{o}\}$
represent valence-bond states where the energy on each X bond has been minimized. That is, a singlet or a triplet-x wavefuntion is placed on each X bond. 
The superscript $e$ denotes `even-triplet' states while $o$ denotes `odd-triplet' states. These states are distinguished by whether the total number of triplet-x's is even or odd. Analogously, $\{y^{e}\}$ and $\{y^{o}\}$ denote states formed by placing singlet or triplet-y states on Y bonds. The former denotes even-triplet states while the latter denotes odd-triplet states. Each family ($\{ x^e\}$, $\{ x^o\}$, $\{ y^e\}$ or $\{ y^o\}$) contains $2^{N/2-1}$ states. 

To determine low-energy eigenstates, we examine two kinds of overlap matrices within this low-energy manifold: (i) Hamiltonian overlap, defined as $H_{jk}=\langle \zeta_j \vert \hat{H} \vert \zeta_k \rangle$, where $\hat{H}$ is the Hamiltonian in the isotropic limit. (ii) Direct overlaps, defined as $O_{jk}=\langle \zeta_j \vert \zeta_k \rangle$. We solve a generalized eigenvalue problem, formulated as 
\begin{eqnarray}
H_{jk}~ v_k = \lambda ~O_{jk}~v_k,
\label{eqn. Gen Eig}
\end{eqnarray}
where $\lambda$ and $v$ represent the eigenvalue and eigenvector respectively. This problem simplifies due to several properties:
(i) Overlaps within each family (e.g., between two states of the $\{ x^{e}\}$ family) are zero. This can be seen from construction. 
(ii) Overlaps between any even-triplet and an odd-triplet state vanish. This can be seen by inserting complete states in the $S^z$-basis to evaluate overlaps. 
(iii) Overlaps between x-odd-triplet states and y-odd-triplet states vanish. See Appendix ~\ref{app. Reduced Subspace} for details. 

On account of these properties, the odd-triplet states remain pinned at a fixed energy. We arrive at a reduced problem with only  $\{ x^{e}\}$ and  $\{ y^{e}\}$. This leads to a generalized eigenvalue problem of dimension $2^{N/2}$. The Hamiltonian and direct overlap matrices have the same form, with details given in Appendix ~\ref{app. Reduced Subspace}. Upon solving the reduced eigenvalue problem, 
we find that the $2^{N/2}$ states of the Hilbert space split into two energy levels with the same degeneracy. The splitting between the two levels decreases with system size. With these calculations, we draw the following conclusions:
\begin{itemize}
\item At the isotropic point for half-integer $S$, we have a ground state degeneracy of $2^{N/2-1}$. The ground states are linear superpositions of x-even-triplet and y-even-triplet states.
\item The first excited state consists of x-odd-triplet and y-odd-triplet states. It has a degeneracy of $2^{N/2}$.
\item For any finite $N$, the ground state degeneracy remains $2^{N/2-1}$ regardless of anisotropy. By adiabatic continuity, the conserved quantities (as defined in Sec.~\ref{sec.model}) in the ground state are the same as in the anisotropic case. We may assign $W$'s independently on each X bond, with a constraint that the product of all $W$'s must be $+1$.
\item In the $N\rightarrow\infty$ thermodynamic limit, the energy separation between the ground state and the first excited states vanishes. 
This will produce a ground state degeneracy of at least $3\times 2^{N/2-1}$.
\end{itemize}

\subsection{Integer $S$}
With periodic boundaries, we obtain a unique ground state for $K_x  >  K_y $ as well as for $K_y  >  K_x$. At the isotropic limit, we consider a low-energy subspace spanned by two states. 
The first is the unique ground state that appears for $K_x >  K_y$. This state is constructed by placing $\vert t_{S,X}\rangle$ on every X bond. The second is the unique ground state for $ K_y  >  K_x$, with $\vert t_{S,Y}\rangle$ placed on every Y bond.

We solve a two-dimensional generalized eigenvalue problem. 
When $\frac{N}{2}$ is even, the ground state is captured by a symmetric combination, and when $\frac{N}{2}$ is odd, the ground state is the asymmetric combination. See Appendix ~\ref{app. Reduced Subspace} for details.  By adiabatic continuity, the conserved quantities are the same as in the anisotropic case. We have $W=+1$ on all bonds.

 \begin{figure}
    \centering
    \includegraphics[width=\columnwidth]{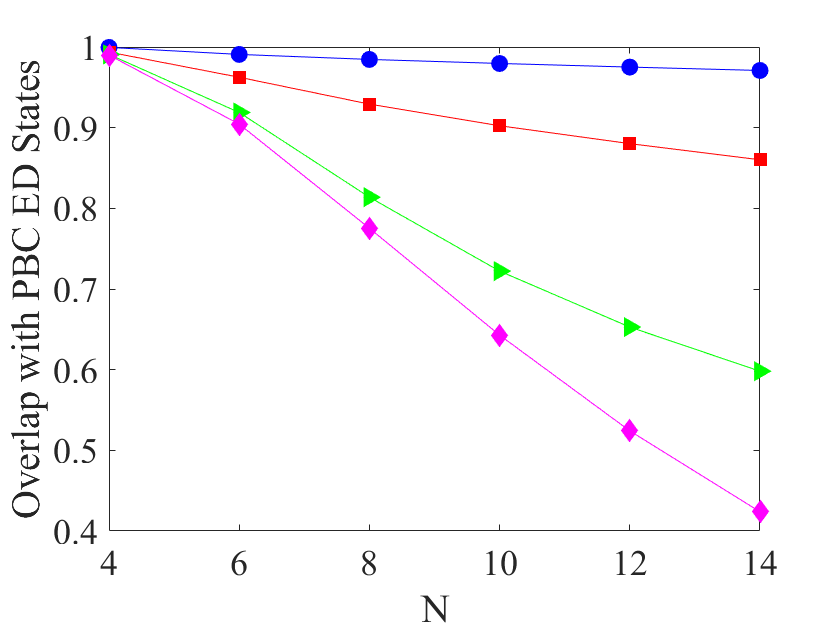}
    \caption{Overlap of proposed ground state for $K_x=K_y$, $S=1$, PBC with the ground state obtained from exact diagonalization. The proposed state is obtained by solving a $2\times 2$ generalized eigenvalue problem. Data for $K_x=K_y=1$ is shown in magenta. We also show data for $K_y/K_x=0.25$ (blue), $K_y/K_x=0.5$ (red) and $K_y/K_x=0.75$ (green). This approach yields a higher overlap than a single valence bond arrangement. }
     \label{fig. Overlap spin one reduced basis }
\end{figure}
\section{Discussion}
\label{sec.discussion}
We provide a valence-bond description for the ground state(s) of the spin-$S$ Kitaev chain. An intriguing feature is the qualitative distinction between integer and half-integer values of $S$. The former yields a unique ground state while the latter exhibits exponential degeneracy. This degeneracy, however, is fragile and easily broken by perturbations. We have explored this for the case of $S=1/2$, where the Jordan-Wigner approach can easily accommodate perturbations such as XY interactions and an external magnetic field along $z$. In both cases, we find that the ground state degeneracy is immediately lost. 
In the case of the isotropic Kitaev chain, using exact diagonalization, we see that adding XY or Heisenberg couplings results in a unique ground state. The dominant weight in this state is captured by a linear superposition of singlet covers on X and Y bonds. Our results could be relevant to materials such as CoNb$_2$O$_6$ \cite{konieczna2025understanding,churchill2024transforming} that are described by a combination of Kitaev and Ising-like couplings. 

An open chain with half-integer $S$ and $K_x>K_y$ effectively hosts Ising variables on each X bond. We have two choices on each bond -- a singlet or a triplet. All Ising configurations produce stable long-lived states within the Kitaev Hamiltonian. 
Based on our perturbative analysis, we expect such states to robust to weak disorder, e.g., when the $K_x$ and $K_y$ coupling are non-uniform. This feature could potentially be useful as a quantum memory. With periodic boundaries, the same system shows topological character,  with a ground state that contains an even number of triplets and an excited state with an odd number of triplets.
It may be possible to store one bit of information in the even/odd character of the wavefunction of the Kitaev chain. 

Kitaev's exact solution for $S=1/2$ moments on a honeycomb lattice has been extended to various two-dimensional and three-dimensional models \cite{mandal2009exactly}. It may be possible to extend our approach for the one-dimensional chain to higher dimensions as well. Our approach and results can be compared with Ref.~\onlinecite{Rousochatzakis2018} which considers semiclassical moments (large $S$) on a honeycomb lattice. Using several layers of arguments, this problem is reduced to defining Ising variables on certain bonds. Naively, this produces an exponential ground state degeneracy. However, perturbative processes couple these variables and give rise to a $\mathbb{Z}_2$ lattice gauge theory. This result is analogous to our result where the global process selects even-triplet states. An exciting future direction is to extend our approach to two dimensions with arbitrary $S$.

\acknowledgments
We thank G. Baskaran, Diptiman Sen and Ajit C. Balram for helpful discussions. 
RG thanks the Office of Global Engagement, IIT Madras for warm hospitality and support. RG acknowledges support from National Sciences and Engineering Research Council of Canada (NSERC) through Discovery Grant 2022-05240. 
RN acknowledges funding from the Center for Quantum Information Theory in Matter and Space-time, IIT Madras, and from the Department of Science and Technology, Government of India, under Grant No. DST/ICPS/QuST/Theme-3/2019/Q69, as well as support from the Mphasis F1 Foundation via the Centre for Quantum Information, Communication, and Computing (CQuICC).

\appendix

\section{Barlow Kitaev chains}
\label{app.Barlow}
We consider a one-dimensional spin chain where every bond has an Ising coupling in the X, Y or Z direction. With $N$ spins, we denote the chain Hamiltonian as $L_1\ldots  L_{N}$, where $L_i =$X, Y or Z. We impose a restriction that adjacent bonds cannot have the same coupling, i.e., $L_i \neq L_{i+1}$. We designate such systems as Barlow Kitaev chains, in analogy with the structure of close-packed solids\cite{KrishnaPandey1981}. Close-packed structures are expressed as a Barlow sequence --- where each entry is A, B or C with the constraint that adjacent entries must be distinct. For example, the face-centred cubic structure is denoted as $\ldots$ABCABC$\ldots$.

We next consider a dual representation which is called the H\"agg code in the context of close-packed structures\cite{Hagg1943}. We denote the `shift' from one-layer to the next as an Ising variable $\sigma_j \sim L_{j+1}-L_j$. A forward shift (X$\rightarrow$Y, Y$\rightarrow$Z or Z$\rightarrow$X) is denoted as $\sigma=+1$. A backward shift (X$\rightarrow$Z, Y$\rightarrow$X or Z$\rightarrow$Y) is represented as $\sigma=-1$. With this scheme, a spin chain with N bonds and periodic boundaries is mapped onto a 1D Ising model with N Ising moments. For example, the original Kitaev chain ($\ldots$XYXYXY$\ldots$) is represented as ($\ldots,+1,-1,+1,-1,\ldots$).

\begin{figure}
\includegraphics[width=0.9\columnwidth]{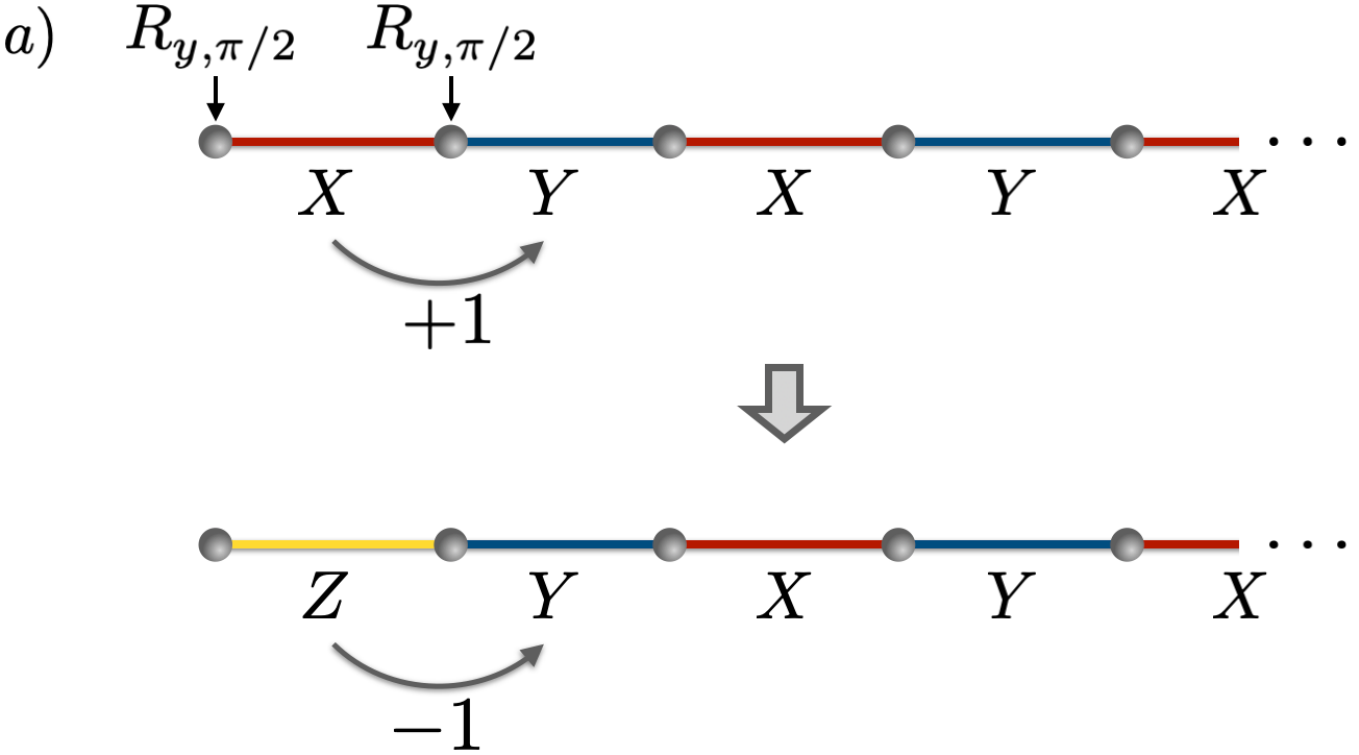}\\
\vspace{0.3in}
\includegraphics[width=\columnwidth]{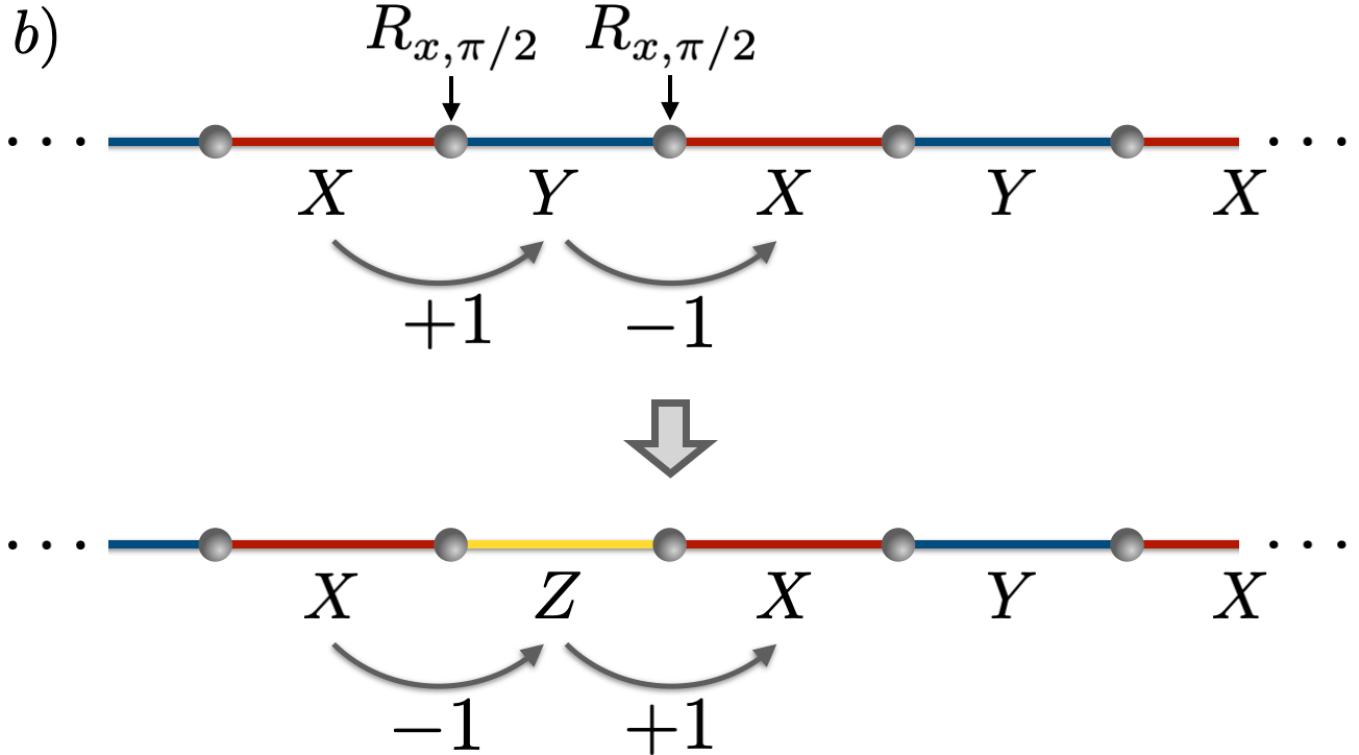}\\
\vspace{0.3in}
\includegraphics[width=\columnwidth]{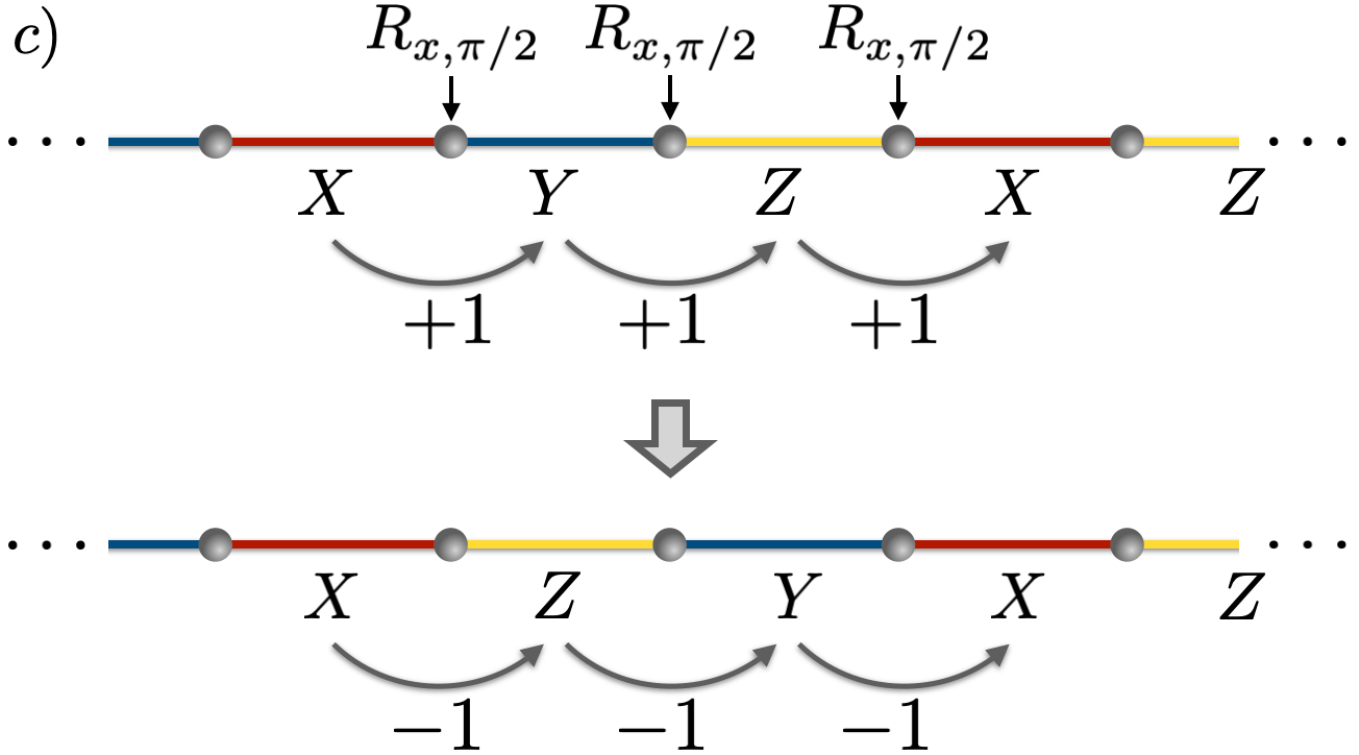}   
    \caption{Mapping one Barlow Kitaev chain to another. (a) With open boundaries, the dangling Ising variable can be flipped by performing a spin rotation at two sites. An example is shown where the dangling edge is converted from X to Z type, by performing spin rotations about Y by $\pi/2$. (b) A pair of adjacent anti-aligned Ising variables can be flipped as shown with an example. (c) Three contiguous and parallel Ising spins can be flipped as shown.    }
\label{fig.Barlow}
\end{figure}

By performing local spin rotations, we can alter the Ising moments in the following ways: (i) with open boundaries, the Ising spin at the boundary can be flipped, (ii) any pair of adjacent anti-aligned Ising moments can be interchanged (e.g., $+1,-1$ $\rightarrow$ $-1,+1$), (iii) any contiguous string of Ising variables that sums to a mutiple of three can be flipped (e.g., $+1,+1,+1$ $\rightarrow$ $-1,-1,-1$). Examples for each of these processes are shown in Fig.~\ref{fig.Barlow}.

Starting from a Barlow Kitaev chain, we can use the three processes described above to alter the sequence of couplings. We argue that any pair of Barlow Kitaev chains can be mapped to one another, as long as they have the same length and the same boundary conditions. 

To demonstrate this, we first define the magnetization of an Ising configuration --- defined as the sum of all Ising variables. Within a fixed magnetization sector, process (ii) suffices to reach every possible Ising configuration. 

If we have source and target configurations (both Barlow Kitaev sequences) that differ in magnetization, process (ii) alone does not suffice. In the case of open boundaries, we can use process (i) to alter the magnetization. With repeated applications of processes (i) and (ii), we can access all values of the magnetization. With periodic boundaries, any valid Barlow Kitaev sequence must necessarily have magnetization that is a multiple of three. This is required to ensure that the bond variables (X/Y/Z) return to the initial value after $N$ shifts. We can then use processes (ii) and (iii) to access any magnetization sector. 

\section{Even-triplet constraint}
\label{app.even}
With half-integer $S$ and periodic boundaries, the ground states have an even-triplet constraint. We have a two-fold choice for the bond-state of each X bond: a singlet or a triplet. This amounts to $2^{N/2}$ possibilities. However, the total number of triplets must be even. Here, we show that this halves the number of possibilities to yield $2^{N/2-1}$ ground states. To see this, consider 
\begin{eqnarray}
 (1-1)^{N/2} &=& \sum_{j} \left(\begin{array}{c}N/2 \\j \end{array}\right) 1^{N/2-j} (-1)^j \\
 &=& N_{j~\mathrm{even}}-N_{j~\mathrm{odd}}. ~~ ~~~~
\end{eqnarray}
As seen from the left hand side, this expression clearly evaluates to zero. On the right hand side, we have used a binomial expansion which yields $2^{N/2}$ terms. 

The index $j$ represents the number of $-1$'s chosen. Each term with even-valued $j$ contributes $+1$ to the sum, while each term with odd valued-$j$ contributes $-1$. This allows us to write the last step, where $N_{j~\mathrm{even}}$ denotes the number ways of choosing an even value of $j$.
We see that $N_{j~\mathrm{even}}=N_{j~\mathrm{odd}}=2^{N/2-1}$.

Here, choosing $(+1)$ and $(-1)$ are analogous to choosing a singlet and a triplet, respectively. We conclude that $N_{\mathrm{even}~\mathrm{triplet}}=N_{\mathrm{odd}~\mathrm{triplet}}=2^{N/2-1}$. That is, the number of even-triplet states (as well as the number of odd-triplet states) is $2^{N/2-1}$.

\section{Details on Reduced Subspace Diagonalisation for Half-Integer and Integer Spin Chains}
\label{app. Reduced Subspace}
\subsection{Half-Integer Spin- Example Spin $\frac{1}{2}$}

By construction, distinct valence bond states on X bonds are mutually orthogonal. Likewise, distinct Y-valence bond states are mutually orthogonal. This follows from the fact that bond-states (singlets or the three triplets) are orthogonal. We next consider overlaps between $x$- and $y$-states. We first consider $S=1/2$ and evaluate $\langle x^o_i \vert y^e_j \rangle$, where $o$ and $e$ represent odd and even triplet numbers respectively. We insert complete states in the $S_z$ basis, 
\begin{equation}
    \langle x^o_i|y_j^e\rangle = \sum_k \langle x^o_i|\{\sigma^z_k \} \rangle \langle \{\sigma^z_k \}|y_j^e\rangle.
\end{equation}
We overlay the $x$- and $y$-states. To have a non-zero contribution, the intermediate state must satisfy the following condition. At the ends of a triplet (singlet) bond, the two $\sigma$'s must be aligned (anti-aligned). With an odd state on $x$-bonds and an even state on $y$-bonds, the total number of triplets is odd. Going around the loop, we have an odd number bonds where the spins are aligned. With such a configuration, we will not return to the initial value of $\sigma_1$ after going around the loop.   
As a result, no intermediate state $\{\sigma^z_k\}$ produces a non-zero overlap. On the same lines, $\langle x^e_i \vert y^o_j \rangle$ also vanishes.

We next consider $ \langle x^o_i|y_j^o\rangle$. We use the property that Pauli matrices square to identity. We have

\begin{equation}
    \langle x^o_i|y_j^o\rangle= \langle x^o_i|\Big(\otimes_{l=1}^N \sigma^y_l\Big) \Big(\otimes_{k=1}^N \sigma^y_k\Big)|y_j^o\rangle=(-1)\langle x^o_i|y_j^o\rangle.
    \label{eq.oddodd}
\end{equation}

Here, we have used two relations: (i) $\otimes_{k=1}^N \sigma^y_k|y_j^o\rangle= (-1)^\frac{N}{2} |y_j^o\rangle$. This relation arises from the fact that both $|t_y\rangle$ and $|s_0\rangle$ states pick up a negative sign upon the action of $\sigma^y$ at both ends of the bond. (ii) $\otimes_{l=1}^N \sigma^y_l|x_j^o\rangle= (-1)^{\frac{N}{2} +1}|x_j^o\rangle$. On an X bond, the action of two $\sigma^y$'s yields a negative sign for $|s\rangle$, but not for $|t_x\rangle$. From Eq.~\ref{eq.oddodd}, we conclude that $\langle x^o_i|y_j^o\rangle=0$.

 We next evaluate overlaps between states of the $\{x^e\}$ and $\{y^e\}$ families. We insert a complete set of states in the $S^z$ basis as before. We follow an argument that is similar to Ref.~\onlinecite{rokhsar1988superconductivity}, but generalized to loops containing an even number of triplets. Only two $\{\sigma^z_k \}$ configurations yield non-zero overlaps. Crucially, both provide contributions that are equal in amplitude and sign. The sign may be positive or negative depending on the precise configuration in $\vert x^e_i\rangle$ and $\vert y_j^o\rangle$.
 In the overlap, every singlet or triplet contributes a factor $\frac{1}{\sqrt{2}}$ (arising from the normalization constant). We obtain 
 \begin{eqnarray}
 \langle x^e_i|y_j^e\rangle=\pm 2\times \left(\frac{ 1}{(\sqrt{2})^\frac{N}{2}}\right)^2=\frac{\pm 1}{2^{\frac{N}{2}-1}}.
 \end{eqnarray}
We consider the reduced subspace with $2^\frac{N}{2}$ states, consisting of $\{x^e\}$ and $\{y^e\}$. We arrive at a generalized eigenvalue problem, with Hamiltonian and overlap matrices given by
\begin{eqnarray}
    H&=&\begin{pmatrix}
        \frac{-N}{8}I && - \frac{N}{4}\mathcal{O} \\\\
        - \frac{N}{4}\mathcal{O}^\dagger && \frac{-N}{8}I
    \end{pmatrix}, \nonumber \\
    O&=&\begin{pmatrix}
        I && \mathcal{O} \\
        \mathcal{O}^\dagger && I
    \end{pmatrix}.
\end{eqnarray}
Here, $I$ is a $2^{\frac{N}{2}-1} \times 2^{\frac{N}{2}-1}$ identity matrix. The matrix $\mathcal{O}$ is of size $2^{\frac{N}{2}-1} \times 2^{\frac{N}{2}-1}$, with every entry being $\mathcal{O}_{ij}=\frac{\pm 1}{2^{\frac{N}{2}-1}}$. The sign depends on the indices $i$ and $j$.

To solve the generalised eigenvalue problem of Eq.~\ref{eqn. Gen Eig}, we note that both $H$ and $O$ have similar matrix structures. We further observe that $\mathcal{O}$ is a normal matrix that can be diagonalized by a unitary transformation $U$ such that $\mathcal{O}_d=U^\dagger \mathcal{O}U$ is diagonal. The eigenvalues of $\mathcal{O}$ are in general complex, they appear as complex conjugate pairs and the absolute value of all eigenvalues is given by $2^{(-)\frac{N-2}{4}}$. We define $W=\begin{pmatrix}
    U && 0\\
    0&& U
\end{pmatrix}$.
In Eq.~\ref{eqn. Gen Eig}, we insert identity in the form of $W^\dagger W$ to obtain $\tilde{H} \tilde{v}=\lambda \tilde{O} \tilde{v}$ where
\begin{eqnarray}
    \tilde{H}&=& W^\dagger H W= \begin{pmatrix}
        \frac{-N}{8}I && - \frac{N}{4}\mathcal{O}_d \\\\
        - \frac{N}{4}\mathcal{O}_d^\dagger && \frac{-N}{8}I
    \end{pmatrix}, \nonumber \\
    \tilde{O}&=& W^\dagger O W=\begin{pmatrix}
        I && \mathcal{O}_d \\
        \mathcal{O}_d^\dagger && I
    \end{pmatrix},~~
    \tilde{v}= W^\dagger v.
\end{eqnarray}
We rearrange the elements to bring $\tilde{H}$ and $\tilde{O}$ to block diagonal form.  

The $i^{\mathrm{th}}$ block is a $2\times 2$ matrix with the form
\begin{eqnarray}
\nonumber    H_{block}&=&\frac{-N}{4}\begin{pmatrix}
        \frac{1}{2} && e_i \\
        e_i^* && \frac{1}{2}
    \end{pmatrix}\\
    &=&\frac{-N}{4}\left(\frac{1}{2}I_{2\times2}+ \mathrm{Re}(e_i)\sigma_x+\mathrm{Im}(e_i)\sigma_y\right);  \\
    O_{block}&=&\begin{pmatrix}
        1 && e_i \\
        e_i^* && 1
    \end{pmatrix}=I_{2\times2}+ \mathrm{Re}(e_i)\sigma_x+\mathrm{Im}(e_i)\sigma_y, \nonumber \\
    |e_i|&=&\frac{1}{2^\frac{N-2}{4}}.
\end{eqnarray}
Here $I_{2\times 2}$ is the identity matrix, $\sigma_{x,y}$ are the Pauli matrices and $e_i,~i=1,...,2^{\frac{N}{2}-1}$ are the eigenvalues of $\mathcal{O}$. With this form, the eigenvalues of $H_{block}$ are seen to be $\frac{-N}{4}(\frac{1}{2}\pm |e_i|)$, while those of $O_{block}$ are $1 \pm |e_i|$. The eigenvectors of both these matrices are identical. This allows the two matrices to be diagonalised by the same unitary transformation. With the $2\times 2$ block structure, we readily  solve the generalised eigenvalue problem. We obtain two eigenvalues, each with a degeneracy $2^{\frac{N}{2}-1}$, given by
\begin{equation}
\lambda=\frac{\frac{-N}{4}(\frac{1}{2}\pm\frac{1}{2^\frac{N-2}{4}})}{1 \pm \frac{1}{2^\frac{N-2}{4}}}.
\end{equation}
The set of $2^\frac{N}{2}$ even-triplet-states splits into two levels. All eigenstates states are linear superpositions of $X$- and $Y$-even triplet states.
The ground states yield substantial overlap with ground state space obtained from exact diagonalization, as seen from Fig.~\ref{fig. Overlap spin half reduced basis}.

Similar arguments can be constructed for arbitrary half-integer $S$. 
\subsection{Integer Spin- Example Spin 1}
Consider a two-dimensional reduced subspace consisting of an X-wavefunction with $\vert t_{S,X}\rangle$ states on all X bonds and a Y-wavefunction with $\vert t_{S,Y}\rangle$ on all Y bonds.

We obtain a Hamiltonian matrix and an overlap matrix of the form
\begin{eqnarray}
    H&=&\begin{pmatrix}
        \frac{-N}{2} && \frac{N}{2^\frac{N}{2}} (-1)^{\frac{N}{2}+1} \\ \\
        \frac{N}{2^{\frac{N}{2}}}(-1)^{\frac{N}{2}+1} && \frac{-N}{2}
    \end{pmatrix}; \nonumber \\
    O&=&\begin{pmatrix}
        1 && \frac{(-1)^\frac{N}{2}}{2^\frac{N}{2}} \\\\
        \frac{(-1)^\frac{N}{2}}{2^\frac{N}{2}} && 1
    \end{pmatrix}.
\end{eqnarray}
Matrix elements are calculated in a manner analogous to the spin $\frac{1}{2}$ case, by inserting a complete set of states in the $S^z$ basis. There is a singlet intermediate state that contributes, with the overlap between X and Y states given by $\frac{1}{2^{\frac{N}{2}}}$. The generalised eigenvalues are given by 
\begin{equation}
    \lambda=\frac{-N(2^{\frac{N}{2}-1}\pm1)}{2^\frac{N}{2}\pm1}.
\end{equation}
The symmetric combination of X and Y states has lower energy when $\frac{N}{2}$ is even, and the asymmetric combination is preferred when $\frac{N}{2}$ is odd. The odd-even effect is due to the difference in signs in the off-diagonal elements of the $H$ and $O$ matrices. The resulting overlap with the exact ground state is plotted in Fig. \ref{fig. Overlap spin one reduced basis }.
\\

 \section{Exact Diagonalization for higher spin}
\begin{figure*}[h]
    \includegraphics[width=0.325\linewidth]{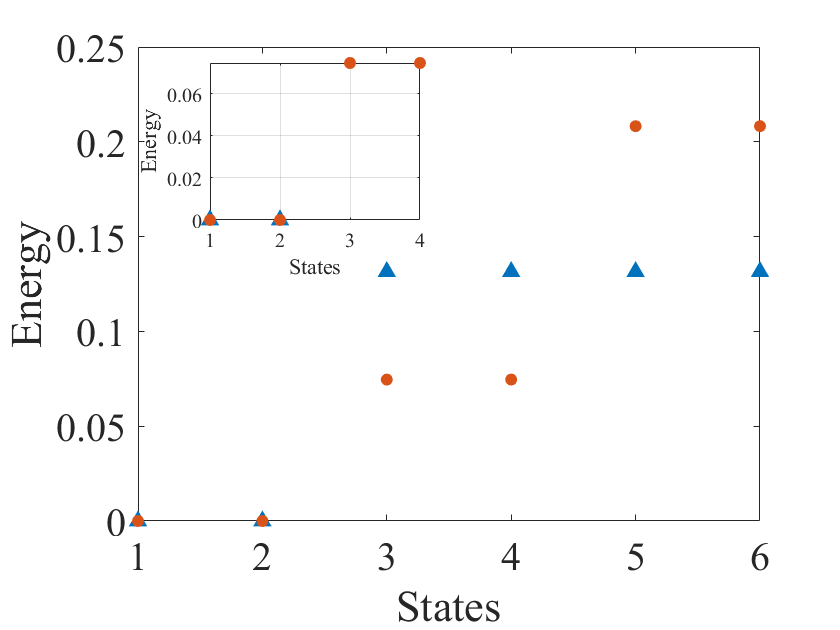}
    \includegraphics[width=0.325\linewidth]{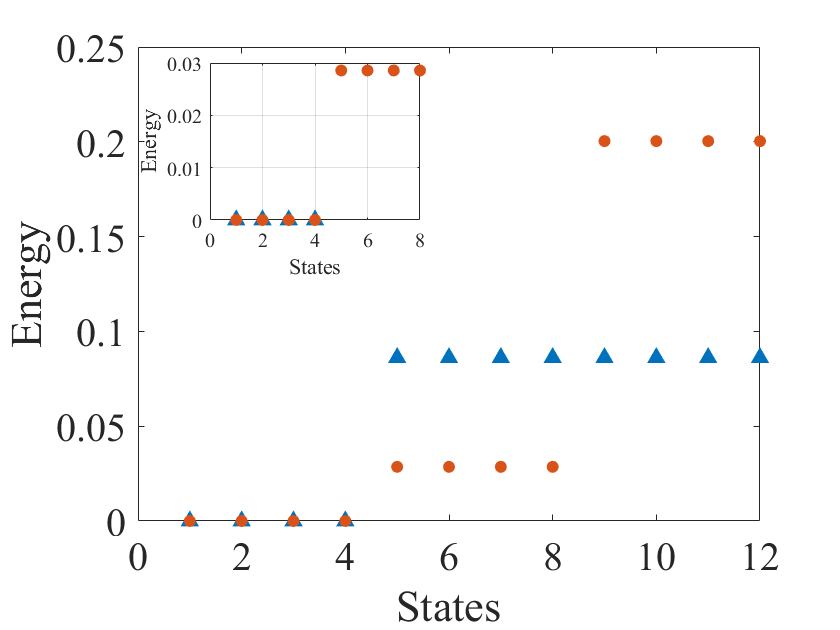}
    \includegraphics[width=0.325\linewidth]{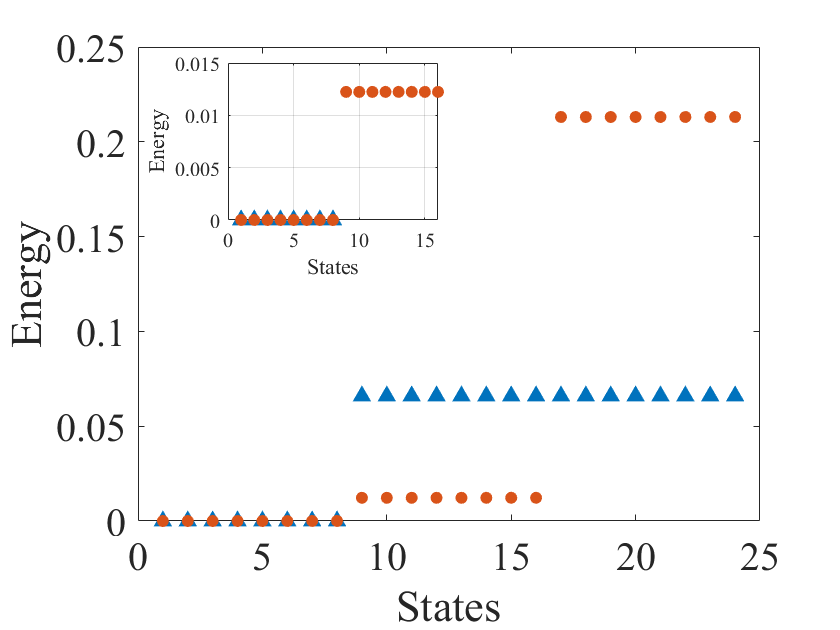}
    \caption{Low-energy spectrum for $S=\frac{3}{2}$ PBC chain for $N=4,6,8$ (from left to right). We only show the lowest few states for clarity. In each case, the ground state has a degeneracy of $2^{\frac{N}{2}-1}$. We fix $K_x=1$. Blue triangles correspond to $K_y=1$, while red circles correspond to $K_y=0.95$. The first excited state is $2^\frac{N}{2}$-fold degenerate at the isotropic point, splitting into two degenerate sets with anisotropy. Energy eigenvalues have been shifted to fix the ground state energy at zero. Insets show the same data, zoomed-in to show the gap between the ground state and the first excitation. }
    \label{fig.3b2}
\end{figure*}

\begin{figure*}[h]
    \includegraphics[width=0.325\linewidth]{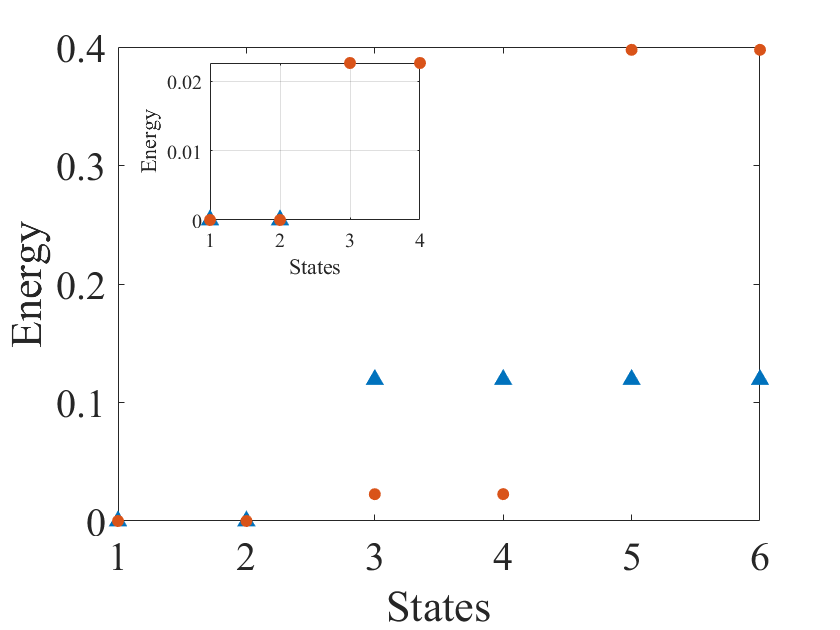}
    \includegraphics[width=0.325\linewidth]{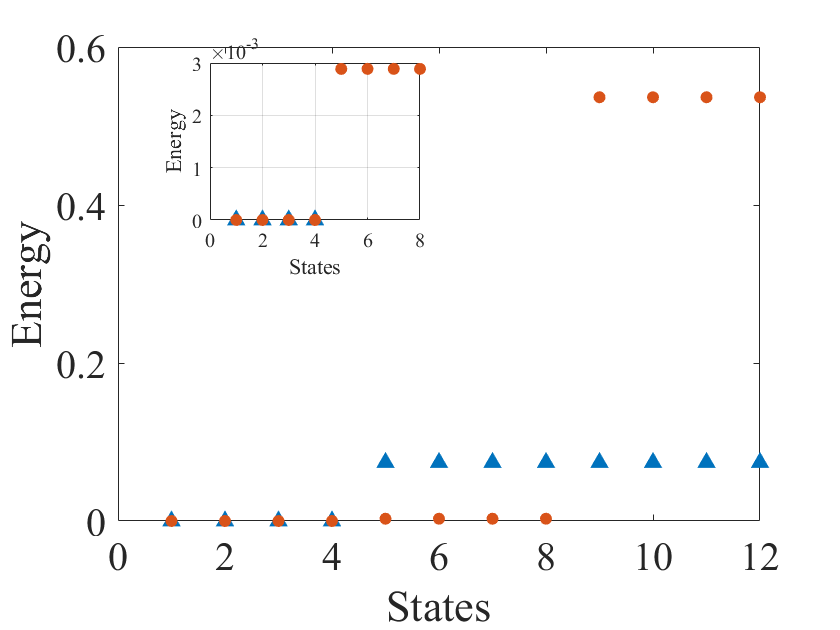}
    \includegraphics[width=0.325\linewidth]{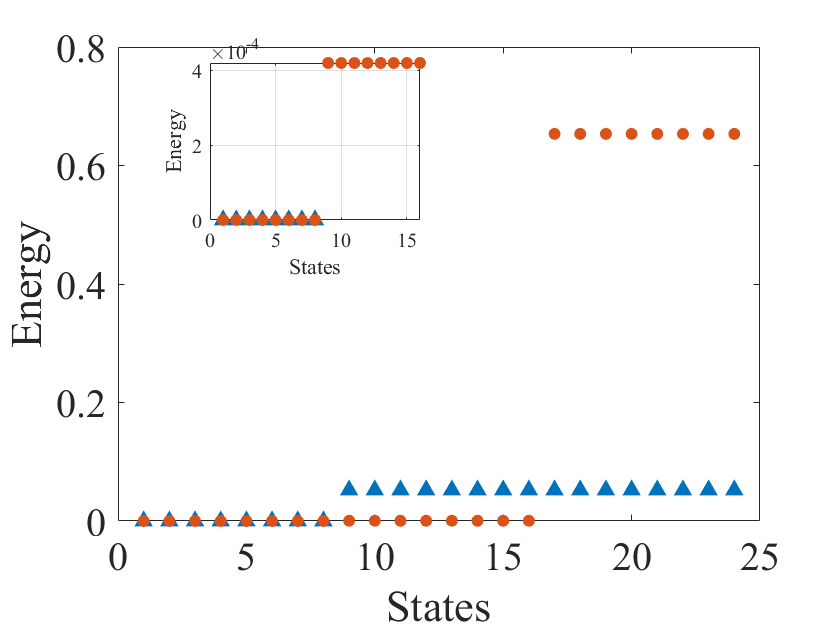}
    \caption{Low-energy spectrum for $S=\frac{5}{2}$ PBC chain for $N=4,6,8$ (from left to right). We only show the lowest few states for clarity. In each case, the ground state has a degeneracy of $2^{\frac{N}{2}-1}$. We fix $K_x=1$. Blue triangles correspond to $K_y=1$, while red circles correspond to $K_y=0.95$. The first excited state is $2^\frac{N}{2}$-fold degenerate at the isotropic point, splitting into two degenerate sets with anisotropy. Energy eigenvalues have been shifted to fix the ground state energy at zero. Insets show the same data, zoomed-in to show the gap between the ground state and the first excitation. }
   \label{fig.5b2}
\end{figure*}

\begin{figure*}
    
    \includegraphics[width=0.45\linewidth]{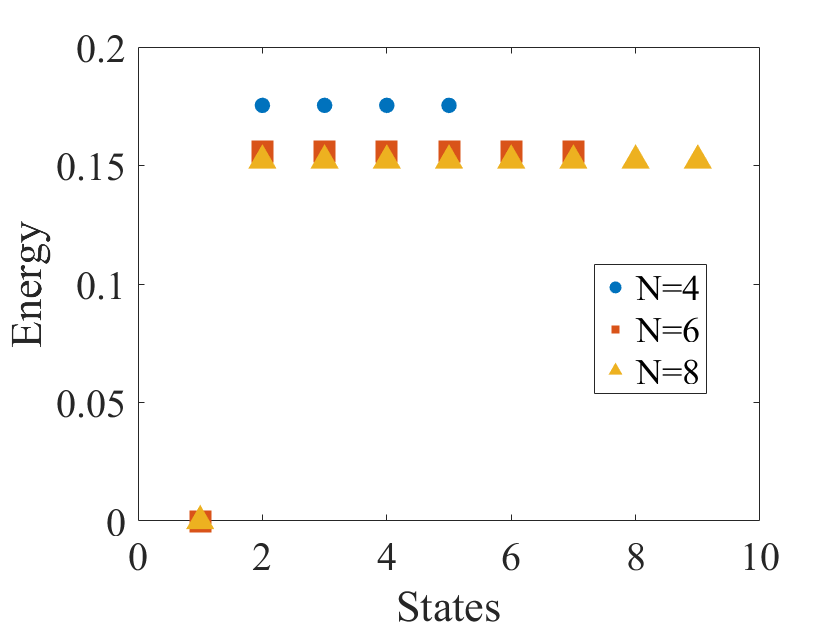}
    \caption {Low-energy spectrum for $S=2$ with $K_x=K_y$ for PBC. Energies are obtained from exact diagonalisation for system sizes $N=4$ to $N=8$. PBC leads to a unique ground state. Y-axis values have been shifted to set the ground state energy to zero. }
    
    \label{fig.S2}
\end{figure*}

We show ED results for $S=\frac{3}{2}, \frac{5}{2}$ in Figs.~\ref{fig.3b2} and \ref{fig.5b2}. The ground-state and first-excited-state degeneracies follow the same pattern as that for $S=\frac{1}{2}$ chains. ED spectra for $S=2$ are shown in Fig.~\ref{fig.S2}. As with the $S=1$ data shown in the main text, we find a unique ground state.

\begin{thebibliography}{28}%
\makeatletter
\providecommand \@ifxundefined [1]{%
 \@ifx{#1\undefined}
}%
\providecommand \@ifnum [1]{%
 \ifnum #1\expandafter \@firstoftwo
 \else \expandafter \@secondoftwo
 \fi
}%
\providecommand \@ifx [1]{%
 \ifx #1\expandafter \@firstoftwo
 \else \expandafter \@secondoftwo
 \fi
}%
\providecommand \natexlab [1]{#1}%
\providecommand \enquote  [1]{``#1''}%
\providecommand \bibnamefont  [1]{#1}%
\providecommand \bibfnamefont [1]{#1}%
\providecommand \citenamefont [1]{#1}%
\providecommand \href@noop [0]{\@secondoftwo}%
\providecommand \href [0]{\begingroup \@sanitize@url \@href}%
\providecommand \@href[1]{\@@startlink{#1}\@@href}%
\providecommand \@@href[1]{\endgroup#1\@@endlink}%
\providecommand \@sanitize@url [0]{\catcode `\\12\catcode `\$12\catcode `\&12\catcode `\#12\catcode `\^12\catcode `\_12\catcode `\%12\relax}%
\providecommand \@@startlink[1]{}%
\providecommand \@@endlink[0]{}%
\providecommand \url  [0]{\begingroup\@sanitize@url \@url }%
\providecommand \@url [1]{\endgroup\@href {#1}{\urlprefix }}%
\providecommand \urlprefix  [0]{URL }%
\providecommand \Eprint [0]{\href }%
\providecommand \doibase [0]{https://doi.org/}%
\providecommand \selectlanguage [0]{\@gobble}%
\providecommand \bibinfo  [0]{\@secondoftwo}%
\providecommand \bibfield  [0]{\@secondoftwo}%
\providecommand \translation [1]{[#1]}%
\providecommand \BibitemOpen [0]{}%
\providecommand \bibitemStop [0]{}%
\providecommand \bibitemNoStop [0]{.\EOS\space}%
\providecommand \EOS [0]{\spacefactor3000\relax}%
\providecommand \BibitemShut  [1]{\csname bibitem#1\endcsname}%
\let\auto@bib@innerbib\@empty
\bibitem [{\citenamefont {Fazekas}(1999)}]{fazekas1999}%
  \BibitemOpen
  \bibfield  {author} {\bibinfo {author} {\bibfnamefont {P.}~\bibnamefont {Fazekas}},\ }\href {https://books.google.co.in/books?id=vLhgDQAAQBAJ} {\emph {\bibinfo {title} {Lecture Notes on Electron Correlation and Magnetism}}},\ EBL-Schweitzer\ (\bibinfo  {publisher} {World Scientific},\ \bibinfo {year} {1999})\BibitemShut {NoStop}%
\bibitem [{\citenamefont {Auerbach}(2012)}]{auerbach2012}%
  \BibitemOpen
  \bibfield  {author} {\bibinfo {author} {\bibfnamefont {A.}~\bibnamefont {Auerbach}},\ }\href {https://books.google.co.in/books?id=d-sHCAAAQBAJ} {\emph {\bibinfo {title} {Interacting Electrons and Quantum Magnetism}}},\ Graduate Texts in Contemporary Physics\ (\bibinfo  {publisher} {Springer New York},\ \bibinfo {year} {2012})\BibitemShut {NoStop}%
\bibitem [{\citenamefont {Kitaev}(2006)}]{kitaev2006anyons}%
  \BibitemOpen
  \bibfield  {author} {\bibinfo {author} {\bibfnamefont {A.}~\bibnamefont {Kitaev}},\ }\bibfield  {title} {\bibinfo {title} {Anyons in an exactly solved model and beyond},\ }\href@noop {} {\bibfield  {journal} {\bibinfo  {journal} {Annals of Physics}\ }\textbf {\bibinfo {volume} {321}},\ \bibinfo {pages} {2} (\bibinfo {year} {2006})}\BibitemShut {NoStop}%
\bibitem [{\citenamefont {Anderson}\ \emph {et~al.}(1987)\citenamefont {Anderson}, \citenamefont {Baskaran}, \citenamefont {Zou},\ and\ \citenamefont {Hsu}}]{anderson1987resonating}%
  \BibitemOpen
  \bibfield  {author} {\bibinfo {author} {\bibfnamefont {P.~W.}\ \bibnamefont {Anderson}}, \bibinfo {author} {\bibfnamefont {G.}~\bibnamefont {Baskaran}}, \bibinfo {author} {\bibfnamefont {Z.}~\bibnamefont {Zou}},\ and\ \bibinfo {author} {\bibfnamefont {T.}~\bibnamefont {Hsu}},\ }\bibfield  {title} {\bibinfo {title} {Resonating--valence-bond theory of phase transitions and superconductivity in {L}a$_2${C}u{O}$_4$-based compounds},\ }\href@noop {} {\bibfield  {journal} {\bibinfo  {journal} {Physical Review Letters}\ }\textbf {\bibinfo {volume} {58}},\ \bibinfo {pages} {2790} (\bibinfo {year} {1987})}\BibitemShut {NoStop}%
\bibitem [{\citenamefont {Liang}\ \emph {et~al.}(1988)\citenamefont {Liang}, \citenamefont {Doucot},\ and\ \citenamefont {Anderson}}]{liang1988some}%
  \BibitemOpen
  \bibfield  {author} {\bibinfo {author} {\bibfnamefont {S.}~\bibnamefont {Liang}}, \bibinfo {author} {\bibfnamefont {B.}~\bibnamefont {Doucot}},\ and\ \bibinfo {author} {\bibfnamefont {P.}~\bibnamefont {Anderson}},\ }\bibfield  {title} {\bibinfo {title} {Some new variational resonating-valence-bond-type wave functions for the spin-$1/2$ antiferromagnetic {H}eisenberg model on a square lattice},\ }\href@noop {} {\bibfield  {journal} {\bibinfo  {journal} {Physical Review Letters}\ }\textbf {\bibinfo {volume} {61}},\ \bibinfo {pages} {365} (\bibinfo {year} {1988})}\BibitemShut {NoStop}%
\bibitem [{\citenamefont {Zhou}\ \emph {et~al.}(2017)\citenamefont {Zhou}, \citenamefont {Kanoda},\ and\ \citenamefont {Ng}}]{zhou2017quantum}%
  \BibitemOpen
  \bibfield  {author} {\bibinfo {author} {\bibfnamefont {Y.}~\bibnamefont {Zhou}}, \bibinfo {author} {\bibfnamefont {K.}~\bibnamefont {Kanoda}},\ and\ \bibinfo {author} {\bibfnamefont {T.-K.}\ \bibnamefont {Ng}},\ }\bibfield  {title} {\bibinfo {title} {Quantum spin liquid states},\ }\href@noop {} {\bibfield  {journal} {\bibinfo  {journal} {Reviews of Modern Physics}\ }\textbf {\bibinfo {volume} {89}},\ \bibinfo {pages} {025003} (\bibinfo {year} {2017})}\BibitemShut {NoStop}%
\bibitem [{\citenamefont {Lee}\ \emph {et~al.}(2006)\citenamefont {Lee}, \citenamefont {Nagaosa},\ and\ \citenamefont {Wen}}]{Lee2006}%
  \BibitemOpen
  \bibfield  {author} {\bibinfo {author} {\bibfnamefont {P.~A.}\ \bibnamefont {Lee}}, \bibinfo {author} {\bibfnamefont {N.}~\bibnamefont {Nagaosa}},\ and\ \bibinfo {author} {\bibfnamefont {X.-G.}\ \bibnamefont {Wen}},\ }\bibfield  {title} {\bibinfo {title} {Doping a mott insulator: Physics of high-temperature superconductivity},\ }\href {https://doi.org/10.1103/RevModPhys.78.17} {\bibfield  {journal} {\bibinfo  {journal} {Rev. Mod. Phys.}\ }\textbf {\bibinfo {volume} {78}},\ \bibinfo {pages} {17} (\bibinfo {year} {2006})}\BibitemShut {NoStop}%
\bibitem [{\citenamefont {Affleck}\ \emph {et~al.}(1988)\citenamefont {Affleck}, \citenamefont {Kennedy}, \citenamefont {Lieb},\ and\ \citenamefont {Tasaki}}]{affleck1988valence}%
  \BibitemOpen
  \bibfield  {author} {\bibinfo {author} {\bibfnamefont {I.}~\bibnamefont {Affleck}}, \bibinfo {author} {\bibfnamefont {T.}~\bibnamefont {Kennedy}}, \bibinfo {author} {\bibfnamefont {E.~H.}\ \bibnamefont {Lieb}},\ and\ \bibinfo {author} {\bibfnamefont {H.}~\bibnamefont {Tasaki}},\ }\bibfield  {title} {\bibinfo {title} {Valence bond ground states in isotropic quantum antiferromagnets},\ }\href@noop {} {\bibfield  {journal} {\bibinfo  {journal} {Communications in Mathematical Physics}\ }\textbf {\bibinfo {volume} {115}},\ \bibinfo {pages} {477} (\bibinfo {year} {1988})}\BibitemShut {NoStop}%
\bibitem [{\citenamefont {Lieb}\ \emph {et~al.}(1961)\citenamefont {Lieb}, \citenamefont {Schultz},\ and\ \citenamefont {Mattis}}]{Lieb1961}%
  \BibitemOpen
  \bibfield  {author} {\bibinfo {author} {\bibfnamefont {E.}~\bibnamefont {Lieb}}, \bibinfo {author} {\bibfnamefont {T.}~\bibnamefont {Schultz}},\ and\ \bibinfo {author} {\bibfnamefont {D.}~\bibnamefont {Mattis}},\ }\bibfield  {title} {\bibinfo {title} {Two soluble models of an antiferromagnetic chain},\ }\href {https://doi.org/https://doi.org/10.1016/0003-4916(61)90115-4} {\bibfield  {journal} {\bibinfo  {journal} {Annals of Physics}\ }\textbf {\bibinfo {volume} {16}},\ \bibinfo {pages} {407} (\bibinfo {year} {1961})}\BibitemShut {NoStop}%
\bibitem [{\citenamefont {Majumdar}\ and\ \citenamefont {Ghosh}(1969)}]{majumdar1969next}%
  \BibitemOpen
  \bibfield  {author} {\bibinfo {author} {\bibfnamefont {C.~K.}\ \bibnamefont {Majumdar}}\ and\ \bibinfo {author} {\bibfnamefont {D.~K.}\ \bibnamefont {Ghosh}},\ }\bibfield  {title} {\bibinfo {title} {On next-nearest-neighbor interaction in linear chain. i},\ }\href@noop {} {\bibfield  {journal} {\bibinfo  {journal} {Journal of Mathematical Physics}\ }\textbf {\bibinfo {volume} {10}},\ \bibinfo {pages} {1388} (\bibinfo {year} {1969})}\BibitemShut {NoStop}%
\bibitem [{\citenamefont {Shastry}(1988)}]{shastry1988exact}%
  \BibitemOpen
  \bibfield  {author} {\bibinfo {author} {\bibfnamefont {B.~S.}\ \bibnamefont {Shastry}},\ }\bibfield  {title} {\bibinfo {title} {Exact solution of an {S}= 1/2 {H}eisenberg antiferromagnetic chain with long-ranged interactions},\ }\href@noop {} {\bibfield  {journal} {\bibinfo  {journal} {Physical Review Letters}\ }\textbf {\bibinfo {volume} {60}},\ \bibinfo {pages} {639} (\bibinfo {year} {1988})}\BibitemShut {NoStop}%
\bibitem [{\citenamefont {Brzezicki}\ \emph {et~al.}(2007)\citenamefont {Brzezicki}, \citenamefont {Dziarmaga},\ and\ \citenamefont {Ole\ifmmode~\acute{s}\else \'{s}\fi{}}}]{Brzezicki2007}%
  \BibitemOpen
  \bibfield  {author} {\bibinfo {author} {\bibfnamefont {W.}~\bibnamefont {Brzezicki}}, \bibinfo {author} {\bibfnamefont {J.}~\bibnamefont {Dziarmaga}},\ and\ \bibinfo {author} {\bibfnamefont {A.~M.}\ \bibnamefont {Ole\ifmmode~\acute{s}\else \'{s}\fi{}}},\ }\bibfield  {title} {\bibinfo {title} {Quantum phase transition in the one-dimensional compass model},\ }\href {https://doi.org/10.1103/PhysRevB.75.134415} {\bibfield  {journal} {\bibinfo  {journal} {Phys. Rev. B}\ }\textbf {\bibinfo {volume} {75}},\ \bibinfo {pages} {134415} (\bibinfo {year} {2007})}\BibitemShut {NoStop}%
\bibitem [{\citenamefont {Nussinov}\ and\ \citenamefont {Van Den~Brink}(2015)}]{nussinov2015compass}%
  \BibitemOpen
  \bibfield  {author} {\bibinfo {author} {\bibfnamefont {Z.}~\bibnamefont {Nussinov}}\ and\ \bibinfo {author} {\bibfnamefont {J.}~\bibnamefont {Van Den~Brink}},\ }\bibfield  {title} {\bibinfo {title} {Compass models: Theory and physical motivations},\ }\href@noop {} {\bibfield  {journal} {\bibinfo  {journal} {Reviews of Modern Physics}\ }\textbf {\bibinfo {volume} {87}},\ \bibinfo {pages} {1} (\bibinfo {year} {2015})}\BibitemShut {NoStop}%
\bibitem [{\citenamefont {Baskaran}\ \emph {et~al.}(2008)\citenamefont {Baskaran}, \citenamefont {Sen},\ and\ \citenamefont {Shankar}}]{baskaran2008spin}%
  \BibitemOpen
  \bibfield  {author} {\bibinfo {author} {\bibfnamefont {G.}~\bibnamefont {Baskaran}}, \bibinfo {author} {\bibfnamefont {D.}~\bibnamefont {Sen}},\ and\ \bibinfo {author} {\bibfnamefont {R.}~\bibnamefont {Shankar}},\ }\bibfield  {title} {\bibinfo {title} {Spin-{S} {K}itaev model: Classical ground states, order from disorder, and exact correlation functions},\ }\href@noop {} {\bibfield  {journal} {\bibinfo  {journal} {Physical Review B—Condensed Matter and Materials Physics}\ }\textbf {\bibinfo {volume} {78}},\ \bibinfo {pages} {115116} (\bibinfo {year} {2008})}\BibitemShut {NoStop}%
\bibitem [{\citenamefont {Khatua}\ \emph {et~al.}(2021)\citenamefont {Khatua}, \citenamefont {Srinivasan},\ and\ \citenamefont {Ganesh}}]{Khatua2021}%
  \BibitemOpen
  \bibfield  {author} {\bibinfo {author} {\bibfnamefont {S.}~\bibnamefont {Khatua}}, \bibinfo {author} {\bibfnamefont {S.}~\bibnamefont {Srinivasan}},\ and\ \bibinfo {author} {\bibfnamefont {R.}~\bibnamefont {Ganesh}},\ }\bibfield  {title} {\bibinfo {title} {State selection in frustrated magnets},\ }\href {https://doi.org/10.1103/PhysRevB.103.174412} {\bibfield  {journal} {\bibinfo  {journal} {Phys. Rev. B}\ }\textbf {\bibinfo {volume} {103}},\ \bibinfo {pages} {174412} (\bibinfo {year} {2021})}\BibitemShut {NoStop}%
\bibitem [{\citenamefont {Gordon}\ and\ \citenamefont {Kee}(2022)}]{gordon2022insights}%
  \BibitemOpen
  \bibfield  {author} {\bibinfo {author} {\bibfnamefont {J.~S.}\ \bibnamefont {Gordon}}\ and\ \bibinfo {author} {\bibfnamefont {H.-Y.}\ \bibnamefont {Kee}},\ }\bibfield  {title} {\bibinfo {title} {Insights into the anisotropic spin-$s$ {K}itaev chain},\ }\href@noop {} {\bibfield  {journal} {\bibinfo  {journal} {Physical Review Research}\ }\textbf {\bibinfo {volume} {4}},\ \bibinfo {pages} {013205} (\bibinfo {year} {2022})}\BibitemShut {NoStop}%
\bibitem [{\citenamefont {Tasaki}(2020)}]{Tasaki2020}%
  \BibitemOpen
  \bibfield  {author} {\bibinfo {author} {\bibfnamefont {H.}~\bibnamefont {Tasaki}},\ }\href {https://doi.org/10.1007/978-3-030-41265-4} {\emph {\bibinfo {title} {Physics and Mathematics of Quantum Many-Body Systems}}},\ Graduate Texts in Physics\ (\bibinfo  {publisher} {Springer, Cham},\ \bibinfo {year} {2020})\ pp.\ \bibinfo {pages} {XVIII + 525},\ \bibinfo {note} {245 b/w illustrations, 2 colour illustrations. Topics: Strongly Correlated Systems, Superconductivity, Mathematical Physics, Statistical Physics and Dynamical Systems, Phase Transitions and Multiphase Systems, Mathematical Methods in Physics}\BibitemShut {NoStop}%
\bibitem [{\citenamefont {Sen}\ \emph {et~al.}(2010)\citenamefont {Sen}, \citenamefont {Shankar}, \citenamefont {Dhar},\ and\ \citenamefont {Ramola}}]{sen2010spin}%
  \BibitemOpen
  \bibfield  {author} {\bibinfo {author} {\bibfnamefont {D.}~\bibnamefont {Sen}}, \bibinfo {author} {\bibfnamefont {R.}~\bibnamefont {Shankar}}, \bibinfo {author} {\bibfnamefont {D.}~\bibnamefont {Dhar}},\ and\ \bibinfo {author} {\bibfnamefont {K.}~\bibnamefont {Ramola}},\ }\bibfield  {title} {\bibinfo {title} {Spin-1 {K}itaev model in one dimension},\ }\href@noop {} {\bibfield  {journal} {\bibinfo  {journal} {Physical Review B—Condensed Matter and Materials Physics}\ }\textbf {\bibinfo {volume} {82}},\ \bibinfo {pages} {195435} (\bibinfo {year} {2010})}\BibitemShut {NoStop}%
\bibitem [{\citenamefont {Yang}\ \emph {et~al.}(2025)\citenamefont {Yang}, \citenamefont {Nocera}, \citenamefont {Xu}, \citenamefont {Ma}, \citenamefont {Adhikary},\ and\ \citenamefont {Affleck}}]{yang2025emergent}%
  \BibitemOpen
  \bibfield  {author} {\bibinfo {author} {\bibfnamefont {W.}~\bibnamefont {Yang}}, \bibinfo {author} {\bibfnamefont {A.}~\bibnamefont {Nocera}}, \bibinfo {author} {\bibfnamefont {C.}~\bibnamefont {Xu}}, \bibinfo {author} {\bibfnamefont {S.}~\bibnamefont {Ma}}, \bibinfo {author} {\bibfnamefont {A.}~\bibnamefont {Adhikary}},\ and\ \bibinfo {author} {\bibfnamefont {I.}~\bibnamefont {Affleck}},\ }\bibfield  {title} {\bibinfo {title} {Emergent {SU}(2)$_1$ conformal symmetry in the spin-{1/2} {K}itaev-{G}amma chain with a {D}zyaloshinskii-{M}oriya interaction},\ }\href@noop {} {\bibfield  {journal} {\bibinfo  {journal} {Physical Review B}\ }\textbf {\bibinfo {volume} {111}},\ \bibinfo {pages} {174414} (\bibinfo {year} {2025})}\BibitemShut {NoStop}%
\bibitem [{\citenamefont {H\"agg}(1943)}]{Hagg1943}%
  \BibitemOpen
  \bibfield  {author} {\bibinfo {author} {\bibfnamefont {G.}~\bibnamefont {H\"agg}},\ }\href@noop {} {\bibfield  {journal} {\bibinfo  {journal} {Ark. Kem. Mineral. Geol.}\ }\textbf {\bibinfo {volume} {163}},\ \bibinfo {pages} {1} (\bibinfo {year} {1943})}\BibitemShut {NoStop}%
\bibitem [{\citenamefont {{Krishna, P.}}\ and\ \citenamefont {{Pandey, D.}}(1981)}]{KrishnaPandey1981}%
  \BibitemOpen
  \bibfield  {author} {\bibinfo {author} {\bibnamefont {{Krishna, P.}}}\ and\ \bibinfo {author} {\bibnamefont {{Pandey, D.}}},\ }\href {https://www.iucr.org/__data/assets/pdf_file/0015/13254/5.pdf} {\bibinfo {title} {Close-packed structures}} (\bibinfo {year} {1981}),\ \bibinfo {note} {{I}nternational {U}nion of {C}rystallography}\BibitemShut {NoStop}%
\bibitem [{\citenamefont {Mohapatra}\ and\ \citenamefont {Balram}(2023)}]{mohapatra2023pronounced}%
  \BibitemOpen
  \bibfield  {author} {\bibinfo {author} {\bibfnamefont {S.}~\bibnamefont {Mohapatra}}\ and\ \bibinfo {author} {\bibfnamefont {A.~C.}\ \bibnamefont {Balram}},\ }\bibfield  {title} {\bibinfo {title} {Pronounced quantum many-body scars in the one-dimensional spin-1 {K}itaev model},\ }\href@noop {} {\bibfield  {journal} {\bibinfo  {journal} {Physical Review B}\ }\textbf {\bibinfo {volume} {107}},\ \bibinfo {pages} {235121} (\bibinfo {year} {2023})}\BibitemShut {NoStop}%
\bibitem [{\citenamefont {You}\ \emph {et~al.}(2022)\citenamefont {You}, \citenamefont {Zhao}, \citenamefont {Ren}, \citenamefont {Sun}, \citenamefont {Li},\ and\ \citenamefont {Ole{\'s}}}]{you2022quantum}%
  \BibitemOpen
  \bibfield  {author} {\bibinfo {author} {\bibfnamefont {W.-L.}\ \bibnamefont {You}}, \bibinfo {author} {\bibfnamefont {Z.}~\bibnamefont {Zhao}}, \bibinfo {author} {\bibfnamefont {J.}~\bibnamefont {Ren}}, \bibinfo {author} {\bibfnamefont {G.}~\bibnamefont {Sun}}, \bibinfo {author} {\bibfnamefont {L.}~\bibnamefont {Li}},\ and\ \bibinfo {author} {\bibfnamefont {A.~M.}\ \bibnamefont {Ole{\'s}}},\ }\bibfield  {title} {\bibinfo {title} {Quantum many-body scars in spin-1 {K}itaev chains},\ }\href@noop {} {\bibfield  {journal} {\bibinfo  {journal} {Physical Review Research}\ }\textbf {\bibinfo {volume} {4}},\ \bibinfo {pages} {013103} (\bibinfo {year} {2022})}\BibitemShut {NoStop}%
\bibitem [{\citenamefont {Konieczna}\ \emph {et~al.}(2025)\citenamefont {Konieczna}, \citenamefont {Kaib}, \citenamefont {Winter},\ and\ \citenamefont {Valent{\'\i}}}]{konieczna2025understanding}%
  \BibitemOpen
  \bibfield  {author} {\bibinfo {author} {\bibfnamefont {A.~A.}\ \bibnamefont {Konieczna}}, \bibinfo {author} {\bibfnamefont {D.~A.}\ \bibnamefont {Kaib}}, \bibinfo {author} {\bibfnamefont {S.~M.}\ \bibnamefont {Winter}},\ and\ \bibinfo {author} {\bibfnamefont {R.}~\bibnamefont {Valent{\'\i}}},\ }\bibfield  {title} {\bibinfo {title} {Understanding the microscopic origin of the magnetic interactions in {C}o{N}b$_2${O}$_6$},\ }\href@noop {} {\bibfield  {journal} {\bibinfo  {journal} {npj Quantum Materials}\ }\textbf {\bibinfo {volume} {10}},\ \bibinfo {pages} {8} (\bibinfo {year} {2025})}\BibitemShut {NoStop}%
\bibitem [{\citenamefont {Churchill}\ and\ \citenamefont {Kee}(2024)}]{churchill2024transforming}%
  \BibitemOpen
  \bibfield  {author} {\bibinfo {author} {\bibfnamefont {D.}~\bibnamefont {Churchill}}\ and\ \bibinfo {author} {\bibfnamefont {H.-Y.}\ \bibnamefont {Kee}},\ }\bibfield  {title} {\bibinfo {title} {Transforming from {K}itaev to disguised {I}sing chain: Application to {C}o{N}b$_2${O}$_6$},\ }\href@noop {} {\bibfield  {journal} {\bibinfo  {journal} {Physical Review Letters}\ }\textbf {\bibinfo {volume} {133}},\ \bibinfo {pages} {056703} (\bibinfo {year} {2024})}\BibitemShut {NoStop}%
\bibitem [{\citenamefont {Mandal}\ and\ \citenamefont {Surendran}(2009)}]{mandal2009exactly}%
  \BibitemOpen
  \bibfield  {author} {\bibinfo {author} {\bibfnamefont {S.}~\bibnamefont {Mandal}}\ and\ \bibinfo {author} {\bibfnamefont {N.}~\bibnamefont {Surendran}},\ }\bibfield  {title} {\bibinfo {title} {Exactly solvable {K}itaev model in three dimensions},\ }\href@noop {} {\bibfield  {journal} {\bibinfo  {journal} {Physical Review B—Condensed Matter and Materials Physics}\ }\textbf {\bibinfo {volume} {79}},\ \bibinfo {pages} {024426} (\bibinfo {year} {2009})}\BibitemShut {NoStop}%
\bibitem [{\citenamefont {Rousochatzakis}\ \emph {et~al.}(2018)\citenamefont {Rousochatzakis}, \citenamefont {Sizyuk},\ and\ \citenamefont {Perkins}}]{Rousochatzakis2018}%
  \BibitemOpen
  \bibfield  {author} {\bibinfo {author} {\bibfnamefont {I.}~\bibnamefont {Rousochatzakis}}, \bibinfo {author} {\bibfnamefont {Y.}~\bibnamefont {Sizyuk}},\ and\ \bibinfo {author} {\bibfnamefont {N.~B.}\ \bibnamefont {Perkins}},\ }\bibfield  {title} {\bibinfo {title} {Quantum spin liquid in the semiclassical regime},\ }\href {https://doi.org/10.1038/s41467-018-03934-1} {\bibfield  {journal} {\bibinfo  {journal} {Nature Communications}\ }\textbf {\bibinfo {volume} {9}},\ \bibinfo {pages} {1575} (\bibinfo {year} {2018})}\BibitemShut {NoStop}%
\bibitem [{\citenamefont {Rokhsar}\ and\ \citenamefont {Kivelson}(1988)}]{rokhsar1988superconductivity}%
  \BibitemOpen
  \bibfield  {author} {\bibinfo {author} {\bibfnamefont {D.~S.}\ \bibnamefont {Rokhsar}}\ and\ \bibinfo {author} {\bibfnamefont {S.~A.}\ \bibnamefont {Kivelson}},\ }\bibfield  {title} {\bibinfo {title} {Superconductivity and the quantum hard-core dimer gas},\ }\href@noop {} {\bibfield  {journal} {\bibinfo  {journal} {Physical Review Letters}\ }\textbf {\bibinfo {volume} {61}},\ \bibinfo {pages} {2376} (\bibinfo {year} {1988})}\BibitemShut {NoStop}%
\end{thebibliography}
%

 \end{document}